\documentclass[aps,pre,final,10pt,onecolumn,superscriptaddress,nofootinbib,showkeys,flushbottom]{revtex4-1}%
\usepackage{amssymb,amsmath,amsfonts}
\usepackage{graphicx,color}
\usepackage[squaren]{SIunits}
\usepackage[]{units}
\usepackage[english]{babel}
\usepackage{tabularx}
\newcolumntype{Y}{>{\centering\arraybackslash}X}

\newcommand{\dif}{\mathrm{d}}%
\newcommand{\Eins}{\mathbf{1}}%
\newcommand{\Diag}{\operatorname{diag}}%

\newcommand{\ii}{\mathrm{i}}%
\newcommand{\Laplace}{\boldsymbol{\triangle}}%
\newcommand{\Nabla}{\vec{\nabla}}%
\newcommand{\pdif}[2]{\frac{\partial#1}{\partial#2}}%
\newcommand{\R}{\mathbb{R}}%
\newcommand{\C}{\mathbb{C}}%
\newcommand{\Real}{\operatorname{Re}}%
\newcommand{\Imag}{\operatorname{Im}}%
\newcommand{\abs}[1]{\lvert#1\rvert}%
\newcommand{\norm}[1]{\lVert#1\rVert}%
\newcommand{\rs}{\vec{r}\hskip1pt'}%

\newcommand{\PP}{\mathcal{P}}%
\newcommand{\QQ}{\mathcal{Q}}%

\begin{document}%
\title{Nonequilibrium dynamics of mixtures of active and passive colloidal particles}%

\author{Raphael Wittkowski}
\email[Electronic address: ]{raphael.wittkowski@uni-muenster.de}
\affiliation{Institut f\"{u}r Theoretische Physik, Westf\"alische Wilhelms-Universit\"{a}t M\"{u}nster, D-48149 M\"{u}nster, Germany}
\affiliation{Center for Nonlinear Science (CeNoS), Westf\"alische Wilhelms-Universit\"{a}t M\"{u}nster, D-48149 M\"{u}nster, Germany}

\author{Joakim Stenhammar} 
\affiliation{Division of Physical Chemistry, Lund University, S-221 00 Lund, Sweden}

\author{Michael E. Cates} 
\affiliation{DAMTP, Centre for Mathematical Sciences, University of Cambridge, Cambridge CB3 0WA, United Kingdom}

\date{\today}

\begin{abstract}
We develop a mesoscopic field theory for the collective nonequilibrium dynamics of multicomponent mixtures of interacting active (i.e., motile) and passive (i.e., nonmotile) colloidal particles with isometric shape in two spatial dimensions. By a stability analysis of the field theory, we obtain equations for the spinodal that describes the onset of a motility-induced instability leading to cluster formation in such mixtures. The prediction for the spinodal is found to be in good agreement with particle-resolved computer simulations. Furthermore, we show that in active-passive mixtures the spinodal instability can be of two different types. One type is associated with a stationary bifurcation and occurs also in one-component active systems, whereas the other type is associated with a Hopf bifurcation and can occur only in active-passive mixtures. Remarkably, the Hopf bifurcation leads to moving clusters. This explains recent results from simulations of active-passive particle mixtures, where moving clusters and interfaces that are not seen in the corresponding one-component systems have been observed. 
\end{abstract}


\maketitle
\newpage

\section{Introduction}
Mixtures of active (i.e., motile) and passive (i.e., nonmotile) colloidal particles exhibit an interesting set of collective behaviors that is quite different from the dynamics of the corresponding one-component systems. Many studies of active-passive mixtures have focused on the dilute regime, where long-range hydrodynamic flows generated by active particles affect the diffusion of passive tracer particles \cite{LeptosGGPG2009,LinTC2011,JepsonMSLMP2013}. At higher densities, where short-range interactions such as excluded-volume interactions become important, other fascinating phenomena arise. Among them are effective depletion-like attractions between passive objects in an active-particle suspension \cite{AngelaniMBRdL2011,NiCSB2015,TanakaLB2017}, crystallization and melting of hard-sphere glasses by doping with active particles \cite{NiCSDB2014,KuemmelSLVB2015}, and mesoscale turbulence mediated by passive particles \cite{HinzPKF2014}. Furthermore, the intriguing phenomenon of motility-induced phase separation (MIPS) \cite{CatesT2015}, whereby purely repulsive active particles spontaneously segregate into dense and dilute phases, has recently been numerically shown to occur also for mixtures of active and passive particles \cite{StenhammarWMC2015,WysockiWG2016}. For purely active systems, where MIPS has already been extensively studied \cite{TailleurC2008,FilyM2012,CatesT2013,BialkeLS2013,ButtinoniBKLBS2013,StenhammarTAMC2013,RednerHB2013,StenhammarMAC2014,SpeckBML2014,WittkowskiTSAMC2014,WysockiWG2014,TakatoriYB2014,SolonSWKKCT2015,TiribocchiWMC2015}, this transition has been theoretically rationalized as a spinodal-like instability occurring for sufficiently large densities and propulsion speeds. A striking feature observed in active-passive mixtures undergoing MIPS is the emergence of persistently moving interfaces and clusters due to a spontaneous breaking of spatial symmetry \cite{StenhammarWMC2015,WysockiWG2016}, which is qualitatively different from the stationary, albeit fluctuating, clusters observed in purely active suspensions. 

To gain a deeper theoretical understanding of MIPS in mixtures of active and passive particles, as well as mixtures of active particles with different properties, we here derive a mesoscopic field theory that describes the collective dynamics of multicomponent mixtures of such particles. In order to restrict the study to the basic features of their nonequilibrium dynamics, we consider unbounded systems in two spatial dimensions and particles with isometric shapes (i.e., disks or spheres in a plane). Furthermore, we neglect hydrodynamic interactions between the particles. The different species therefore differ only in their radii and motilities. By using appropriate approximations for the full (multidimensional) pair-correlation functions of the particles, we derive spinodal criteria based on coefficients that can be readily evaluated from computer simulations in the one-phase parameter regime. The predicted spinodals are in good accordance with the instability region observed in simulations of active-passive mixtures. We furthermore show how two different types of instabilities arise, depending on the properties of the mixtures: a stationary bifurcation, corresponding to MIPS in one-component active systems, and a Hopf bifurcation, which can only occur in mixtures and is associated with a steady state with moving clusters. This distinction is likely to be responsible for the intriguing collective dynamics observed in computer simulations of active-passive mixtures undergoing MIPS \cite{StenhammarWMC2015,WysockiWG2016}.

The article in organized as follows: in Sec.\ \ref{sec:equations} we derive dynamic equations for multicomponent mixtures of particles with different motilities as well as useful approximations of these equations. Based on these, in Sec.\ \ref{sec:instability} we address the stability of mixtures of active and passive particles and derive a condition for the onset of a dynamical instability in such mixtures. The good agreement between our analytic results and particle-resolved computer simulations of mixtures of active and passive particles is demonstrated in Sec.\ \ref{sec:simulations}. We finally conclude in Sec.\ \ref{sec:conclusions}.

\section{\label{sec:equations}Dynamic equations for mixtures}
We consider an $n$-component mixture of colloidal particles, each with different but constant magnitudes of self-propulsion. In addition to the deterministic self-propulsion, the positions and orientations of the particles (so-called ``active Brownian particles'' \cite{BechingerdLLRVV2016}) undergo translational and rotational diffusion, respectively. 
In the following, $n$ denotes the number of different species of active or passive colloidal particles and $N_{\mu}$ is the total number of particles of species $\mu\in\{1,\dotsc,n\}$. 
The translational and rotational diffusion coefficients of the particles are $D^{\mu}_{\mathrm{T}}$ and $D^{\mu}_{\mathrm{R}}$, respectively, and the particles' motilities are set by active driving forces $F^{\mu}_{\mathrm{A}}\geqslant 0$ \cite{tenHagenWTKBL2015}, which are zero for passive particles. $\vec{r}^{\mu}_{i}(t)$ is the position and $\phi^{\mu}_{i}(t)$ is the orientation of the $i$th particle of species $\mu$ at time $t$. The interactions between a particle of species $\mu$ and a particle of species $\nu$ are described by the pair-interaction potentials 
$U^{(\mu\nu)}_{2}(\norm{\vec{r}^{\mu}_{i}\!-\vec{r}^{\nu}_{j}})$, where $U^{(\mu\nu)}_{2}(r)$ and $U^{(\nu\mu)}_{2}(r)$ can be different\footnote{This is the case especially for colloidal particles that catalyze chemical reactions on their surfaces and breaks the action-reaction symmetry stated by Newton's third law. As an interesting consequence of such asymmetric interactions, anisotropic clusters of \textit{passive} particles can become self-propelled \cite{SotoG2014}. However, in this article we do not further study such systems. Our simulations described in Sec.\ \ref{sec:simulations} use the same interaction potential for all particles.} and $\norm{\,\cdot\,}$ is the Euclidean norm.

\subsection{Exact dynamic equations}
The microscopic active Brownian motion of an $n$-component mixture of these colloidal particles with different constant motilities is described by the 
\textit{Langevin equations}\footnote{In the limit of a one-component system, these Langevin equations reduce to the simpler ones that have been used in Refs.\ \cite{RednerHB2013,BialkeLS2013}.} 
{\allowdisplaybreaks
\begin{align}%
\begin{split}%
\dot{\vec{r}}^{\mu}_{i}(t)&=\beta D^{\mu}_{\mathrm{T}}\Big(\!\vec{F}^{\mu}_{\mathrm{int},i}(\{\vec{r}^{\mu}_{i}\}) + F^{\mu}_{\mathrm{A}}\hat{u}(\phi^{\mu}_{i})\Big) 
+ \vec{\xi}^{\mu}_{\mathrm{T},i}(t) \;, 
\label{eq:LangevinI}%
\end{split}\\%
\begin{split}%
\dot{\phi}^{\mu}_{i}(t)&=\xi^{\mu}_{\mathrm{R},i}(t) 
\label{eq:LangevinII}%
\end{split}%
\end{align}}%
with the interaction forces 
\begin{equation}
\begin{split}%
\vec{F}^{\mu}_{\mathrm{int},i}(\{\vec{r}^{\mu}_{i}\})
=-\underset{(\mu,i)\neq(\nu,j)}{\sum^{n}_{\nu=1}\sum^{N_{\nu}}_{j=1}}  
\Nabla_{\vec{r}^{\mu}_{i}} U^{(\mu\nu)}_{2}(\norm{\vec{r}^{\mu}_{i}\!-\vec{r}^{\nu}_{j}}) \;.
\end{split}%
\end{equation}
Here, an overdot denotes differentiation with respect to time and $\beta=1/(k_{\mathrm{B}}T)$ is the inverse thermal energy with the Boltzmann constant $k_{\mathrm{B}}$ and the absolute temperature $T$ of the implicit solvent.
Furthermore, $\hat{u}(\phi)=(\cos(\phi),\sin(\phi))^{\mathrm{T}}$ is a normalized orientation vector and $\Nabla_{\vec{r}^{\mu}_{i}}=(\partial_{x^{\mu}_{i}},\partial_{y^{\mu}_{i}})$ is the del symbol that involves partial derivatives with respect to the elements of the vector $\vec{r}^{\mu}_{i}=(x^{\mu}_{i},y^{\mu}_{i})^{\mathrm{T}}$. 
The elements of the translational noise $\vec{\xi}^{\mu}_{\mathrm{T},i}(t)$ and the rotational noise $\xi^{\mu}_{\mathrm{R},i}(t)$ in the Langevin equations are statistically independent Gaussian white noises with mean values
$\langle\vec{\xi}^{\mu}_{\mathrm{T},i}(t)\rangle=\vec{0}$ and $\langle\xi^{\mu}_{\mathrm{R},i}(t)\rangle=0$, 
and correlations 
$\langle\vec{\xi}^{\mu}_{\mathrm{T},i}(t_{1})\!\otimes\!\vec{\xi}^{\nu}_{\mathrm{T},j}(t_{2})\rangle =2D^{\mu}_{\mathrm{T}}\Eins \delta_{ij}\delta_{\mu\nu}\delta(t_{1}-t_{2})$
and 
$\langle\xi^{\mu}_{\mathrm{R},i}(t_{1})\xi^{\nu}_{\mathrm{R},j}(t_{2})\rangle=2D^{\mu}_{\mathrm{R}} \delta_{ij}\delta_{\mu\nu}\delta(t_{1}-t_{2})$, where $\otimes$ denotes the dyadic product and $\Eins$ is the identity matrix.

The collective dynamics of the $n$-component mixture can equivalently be described by the \textit{Smoluchowski equation} (see Refs.\ \cite{Dhont1996,Risken1996} for details) 
\begin{equation}%
\begin{split}%
\dot{P}(\{\vec{r}^{\mu}_{i}\},\{\phi^{\mu}_{i}\},t)&=\sum^{n}_{\mu=1}\sum^{N_{\mu}}_{i=1}
D^{\mu}_{\mathrm{T}}\Nabla_{\vec{r}^{\mu}_{i}}\!\cdot\!\Big(
\Nabla_{\vec{r}^{\mu}_{i}}P(\{\vec{r}^{\mu}_{i}\},\{\phi^{\mu}_{i}\},t)
-\beta\big(\vec{F}^{\mu}_{\mathrm{int},i}(\{\vec{r}^{\mu}_{i}\})
+F^{\mu}_{\mathrm{A}}\hat{u}(\phi^{\mu}_{i})\big)
P(\{\vec{r}^{\mu}_{i}\},\{\phi^{\mu}_{i}\},t)\!\Big)\\
&\quad\:\!+\sum^{n}_{\mu=1}\sum^{N_{\mu}}_{i=1}D^{\mu}_{\mathrm{R}}
\partial^{2}_{\phi^{\mu}_{i}}P(\{\vec{r}^{\mu}_{i}\},\{\phi^{\mu}_{i}\},t) 
\end{split}%
\label{eq:SG}%
\end{equation}%
with the many-particle probability density $P(\{\vec{r}^{\mu}_{i}\},\{\phi^{\mu}_{i}\},t)$. Here, the operator $\partial_{\phi^{\mu}_{i}}$ means partial differentiation with respect to $\phi^{\mu}_{i}$. 
By integrating away all degrees of freedom except for $\vec{r}^{\mu}_{1}$ and $\phi^{\mu}_{1}$, 
the one-particle densities \cite{WittkowskiL2011}
\begin{equation}
\begin{split}%
\rho_{\mu}(\vec{r}^{\mu}_{1},\phi^{\mu}_{1},t)&=N_{\mu} \Bigg(
\!\underset{(\mu,1)\neq(\nu,j)}{\prod^{n}_{\nu=1}\prod^{N_{\nu}}_{j=1}}\!\!
\int_{\R^{2}}\!\!\!\!\!\:\!\dif^{2}r^{\nu}_{j}\!\int^{2\pi}_{0}\!\!\!\!\!\!\!\dif\phi^{\nu}_{j}\Bigg)
P(\{\vec{r}^{\mu}_{i}\},\{\phi^{\mu}_{i}\},t)
\end{split}%
\label{eq:integration}%
\end{equation}
can be obtained from $P(\{\vec{r}^{\mu}_{i}\},\{\phi^{\mu}_{i}\},t)$. 
Many-particle densities like the two-particle densities $\rho^{(2)}_{\mu\nu}(\vec{r},\rs\!,\phi,\phi'\!,t)$ can be obtained analogously \cite{WittkowskiLB2012}. 
Applying the integration \eqref{eq:integration} to the Smoluchowski equation \eqref{eq:SG} and renaming $\vec{r}^{\mu}_{1}\to\vec{r}=(x,y)^{\mathrm{T}}$ and $\phi^{\mu}_{1}\to\phi$ leads to the dynamic equations 
\begin{equation}
\begin{split}%
\dot{\rho}_{\mu}(\vec{r},\phi,t)&=D^{\mu}_{\mathrm{T}}\Laplace_{\vec{r}}\rho_{\mu}(\vec{r},\phi,t)
+D^{\mu}_{\mathrm{R}}\partial^{2}_{\phi}\rho_{\mu}(\vec{r},\phi,t) 
-v^{\mu}_{0}\Nabla_{\vec{r}}\!\cdot\!\big(\hat{u}(\phi)\rho_{\mu}(\vec{r},\phi,t)\big) \\
&\quad\:\!+\beta D^{\mu}_{\mathrm{T}} \sum^{n}_{\nu=1}\Nabla_{\vec{r}}\!\cdot\!\!\!\int_{\R^{2}}\!\!\!\!\!\:\!\dif^{2}r'\,
U^{(\mu\nu)\prime}_{2}(\norm{\vec{r}-\rs}) \frac{\vec{r}-\rs}{\norm{\vec{r}-\rs}}
\int^{2\pi}_{0}\!\!\!\!\!\!\!\dif\phi'\,\rho^{(2)}_{\mu\nu}(\vec{r},\rs\!,\phi,\phi'\!,t) 
\end{split}%
\label{eq:dynamicalI}%
\end{equation}
for the one-particle densities $\rho_{\mu}(\vec{r},\phi,t)$. Here, $v^{\mu}_{0}=\beta D^{\mu}_{\mathrm{T}}F^{\mu}_{\mathrm{A}}$ denotes the speed of a free particle of species $\mu$ and we used the notation $U^{(\mu\nu)\prime}_{2}(r)=\dif U^{(\mu\nu)}_{2}(r)/\dif r$. 
The dynamics of the one-particle densities $\rho_{\mu}(\vec{r},\phi,t)$ depends on the two-particle densities $\rho^{(2)}_{\mu\nu}(\vec{r},\rs\!,\phi,\phi'\!,t)$, which can be decomposed into the one-particle densities $\rho_{\mu}(\vec{r},\phi,t)$ and the pair-correlation functions $g_{\mu\nu}(\vec{r},\rs\!,\phi,\phi'\!,t)$:
\begin{equation}
\begin{split}%
\rho^{(2)}_{\mu\nu}(\vec{r},\rs\!,\phi,\phi'\!,t)&=
\rho_{\mu}(\vec{r},\phi,t) \rho_{\nu}(\rs\!,\phi'\!,t) g_{\mu\nu}(\vec{r},\rs\!,\phi,\phi'\!,t) \;. 
\end{split}%
\label{eq:rhoII}%
\end{equation}
Notice that this standard definition of $g_{\mu\nu}(\vec{r},\rs\!,\phi,\phi'\!,t)$ implies that
\begin{equation}
\lim_{\norm{\vec{r}-\rs}\to\infty} g_{\mu\nu}(\vec{r},\rs\!,\phi,\phi'\!,t) = 1 - \frac{\delta_{\mu\nu}}{N_{\mu}} \;,
\end{equation}
where the term $\delta_{\mu\nu}/N_{\mu}$ is usually negligibly small.
As a further consequence of Eq.\ \eqref{eq:rhoII}, the conventional radial distribution functions cannot be simply obtained by an angular integration and a time average of $g_{\mu\nu}(\vec{r},\rs\!,\phi,\phi'\!,t)$. Instead, they need to be calculated from $\rho^{(2)}_{\mu\nu}(\vec{r},\rs\!,\phi,\phi'\!,t)$.

Inserting Eq.\ \eqref{eq:rhoII} into Eq.\ \eqref{eq:dynamicalI} results in the exact \textit{dynamic equations for the one-particle densities}
\begin{equation}
\begin{split}%
\dot{\rho}_{\mu}(\vec{r},\phi,t)&=D^{\mu}_{\mathrm{T}}\Laplace_{\vec{r}}\rho_{\mu}(\vec{r},\phi,t)
+D^{\mu}_{\mathrm{R}}\partial^{2}_{\phi}\rho_{\mu}(\vec{r},\phi,t) 
-v^{\mu}_{0}\Nabla_{\vec{r}}\!\cdot\!\big(\hat{u}(\phi)\rho_{\mu}(\vec{r},\phi,t)\big) \\
&\quad\:\!+\beta D^{\mu}_{\mathrm{T}} \sum^{n}_{\nu=1}\Nabla_{\vec{r}}\!\cdot\!\bigg(
\rho_{\mu}(\vec{r},\phi,t)\!\int_{\R^{2}}\!\!\!\!\!\:\!\dif^{2}r'\,
U^{(\mu\nu)\prime}_{2}(\norm{\vec{r}-\rs})\frac{\vec{r}-\rs}{\norm{\vec{r}-\rs}}\!\:\!
\int^{2\pi}_{0}\!\!\!\!\!\!\!\dif\phi'\, 
g_{\mu\nu}(\vec{r},\rs\!,\phi,\phi'\!,t) \rho_{\nu}(\rs\!,\phi'\!,t)\!\bigg) \:\!.
\end{split}%
\label{eq:dynamicalII}%
\end{equation}
The pair-correlation functions $g_{\mu\nu}(\vec{r},\rs\!,\phi,\phi'\!,t)$, however, are unknown and have to be approximated appropriately.

\subsection{\label{sec:Approximation}Approximate dynamic equations}
\begin{figure}[ht]
\includegraphics[width=0.35\linewidth]{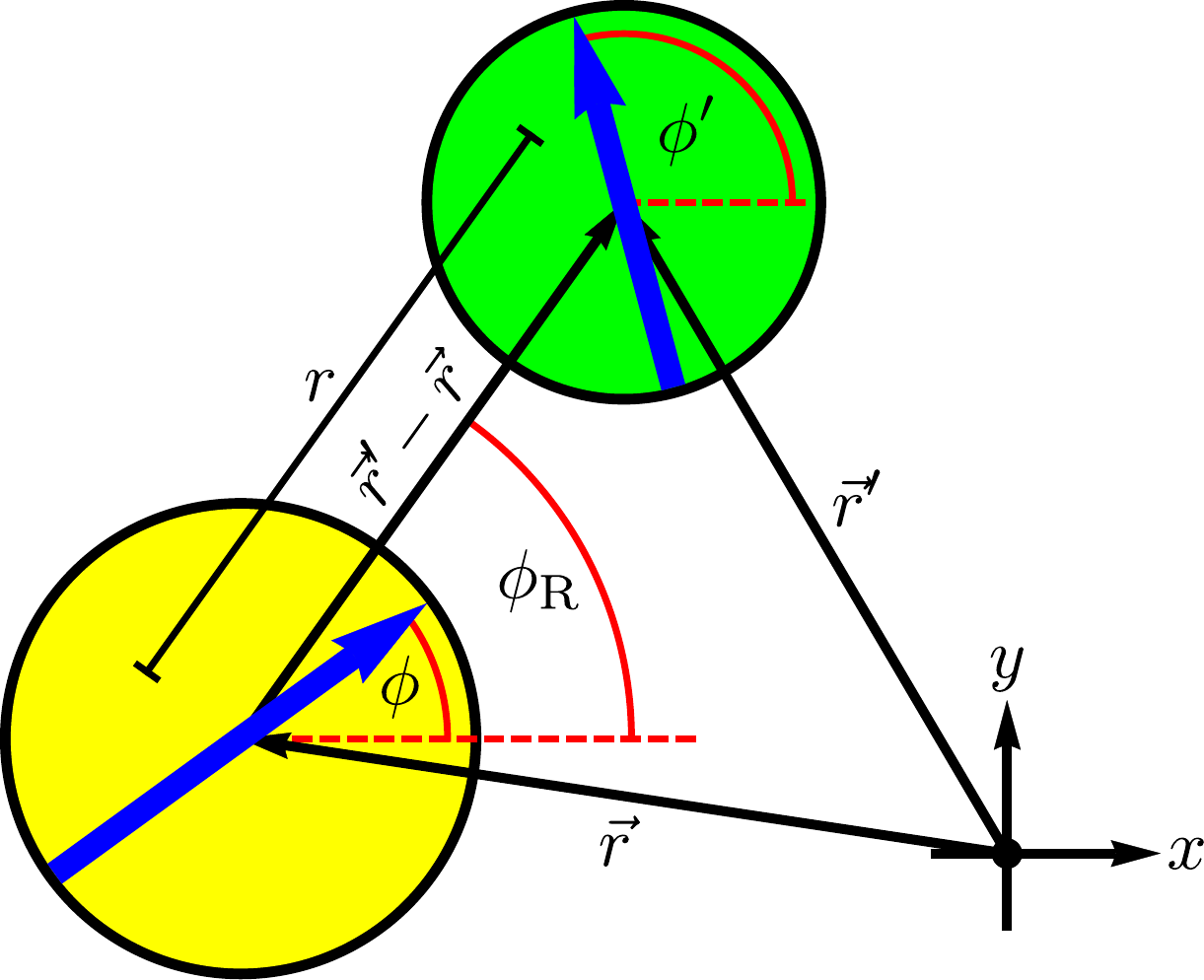}%
\caption{\label{fig:Koordinaten}Absolute and relative positions and orientations of two different isometric colloidal particles}%
\end{figure}

In order to approximate the pair-correlation functions, we first assume global translational invariance of the system, leading to $g_{\mu\nu}(\vec{r},\rs\!,\phi,\phi'\!,t) = g_{\mu\nu}(\rs\!-\vec{r},\phi,\phi'\!,t)$. With the parametrization $\rs-\vec{r}=r\hat{u}(\phi_{\mathrm{R}})$ (see Fig.\ \ref{fig:Koordinaten}) this can be rewritten as $g_{\mu\nu}(r,\phi_{\mathrm{R}},\phi,\phi'\!,t)$.
Assuming also invariance of the system with respect to global rigid rotations, the number of variables of the pair-correlation functions can be reduced further, yielding $g_{\mu\nu}(r,\phi_{\mathrm{R}}-\phi,\phi'\!-\phi,t)$. In addition, we assume that the pair-correlation functions are not explicitly time-dependent: $g_{\mu\nu}(r,\phi_{\mathrm{R}}-\phi,\phi'\!-\phi)$.
These common approximations become exact when the system is in a homogeneous steady state.  
In accordance with Ref.\ \cite{BialkeLS2013}, we now further approximate $g_{\mu\nu}(r,\phi_{\mathrm{R}}-\phi,\phi'\!-\phi)$ by $g_{\mu\nu}(r,\phi_{\mathrm{R}}-\phi)$ with the symmetry property $g_{\mu\nu}(r,-\theta)=g_{\mu\nu}(r,\theta)$. Notice that the pair-correlation functions $g_{\mu\nu}(r,\phi_{\mathrm{R}}-\phi)$ still depend on the driving forces $F^{\mu}_{\mathrm{A}}$ of the active particles and on the constant average particle number densities
$\bar{\rho}_{\mu}=\langle\rho_{\mu}(\vec{r},t)\rangle_{\vec{r}}$, where $\rho_{\mu}(\vec{r},t)=\int^{2\pi}_{0}\!\!\dif\phi\,\rho_{\mu}(\vec{r},\phi,t)$ are the orientation-independent translational densities (i.e., concentration fields) and $\langle\,\cdot\,\rangle_{\vec{r}}$ denotes a spatial average.
Regarding the pair-interaction potentials $U^{(\mu\nu)}_{2}(r)$ we assume short-range interactions between the particles. 
This allows us to simplify the final term of Eq.\ \eqref{eq:dynamicalII} through a gradient expansion \cite{YangFG1976,Evans1979,EmmerichEtAl2012}. 

With these approximations, Eq.\ \eqref{eq:dynamicalII} simplifies to the 
\textit{dynamic equations for $n$-component mixtures of active and passive particles}
\begin{equation}
\begin{split}%
\dot{\rho}_{\mu}(\vec{r},\phi,t)&=D^{\mu}_{\mathrm{T}}\Laplace_{\vec{r}}\rho_{\mu}(\vec{r},\phi,t)
+D^{\mu}_{\mathrm{R}}\partial^{2}_{\phi}\rho_{\mu}(\vec{r},\phi,t) 
-\Nabla_{\vec{r}}\!\cdot\!\big(v_{\mu}(\vec{r},\phi,t) \hat{u}(\phi)\rho_{\mu}(\vec{r},\phi,t)\big) \\
&\quad\:\! +\beta D^{\mu}_{\mathrm{T}} \Nabla_{\vec{r}}\!\cdot\!\big(
\rho_{\mu}(\vec{r},\phi,t)\Nabla_{\vec{r}} \Upsilon_{\mu}(\vec{r},\phi,t) \big)
\end{split}%
\label{eq:DynamischeGlg}%
\end{equation}
with the effective speeds of the active particles 
\begin{equation}
\begin{split}%
v_{\mu}(\vec{r},\phi,t) = v^{\mu}_{0} -\beta D^{\mu}_{\mathrm{T}} F^{\mu}_{\mathrm{res}}(\vec{r},\phi,t) 
\end{split}%
\label{eq:v}%
\end{equation}
that describe the slow-down of active particles due to collisions \cite{CatesT2013,BialkeLS2013}.
Here, the resistive forces 
\begin{equation}
\begin{split}%
F^{\mu}_{\mathrm{res}}(\vec{r},\phi,t)&=\sum^{n}_{\nu=1} \sum^{\infty}_{m=1}
\!\!\!\sum^{\lfloor\frac{m-1}{2}\rfloor}_{k=0}\!\!\sum^{k}_{l=0}  
\bigg(\frac{A^{(\mu\nu)}_{m,2k}}{(m-2k)!(2k)!} -\frac{A^{(\mu\nu)}_{m,2k+2}}{(m-2k-1)!(2k+1)!}\bigg) 
\frac{(-1)^{l}k!}{(k-l)!l!} \Laplace_{\vec{r}}^{k-l} \big(\hat{u}(\phi)\!\cdot\!\Nabla_{\vec{r}}\big)^{m-2(k-l)} 
\rho_{\nu}(\vec{r},t) \\
&\quad\:\!+\sum^{n}_{\nu=1}\sum^{\infty}_{m=0}\sum^{m}_{l=0} \frac{A^{(\mu\nu)}_{2m,2m}}{(2m)!}\frac{(-1)^{l}m!}{(m-l)!l!}
\Laplace_{\vec{r}}^{m-l} \big(\hat{u}(\phi)\!\cdot\!\Nabla_{\vec{r}}\big)^{2l} \rho_{\nu}(\vec{r},t)  
\end{split}\raisetag{3.7em}%
\label{eq:Fres}%
\end{equation}
and the functions 
\begin{equation}
\begin{split}%
\Upsilon_{\mu}(\vec{r},\phi,t)&=\sum^{n}_{\nu=1} \sum^{\infty}_{m=1}
\!\!\!\sum^{\lfloor\frac{m-1}{2}\rfloor}_{k=0}\!\!\sum^{k}_{l=0} 
\frac{A^{(\mu\nu)}_{m,2k+2}}{(m-2k-1)!(2k+1)!} 
\frac{(-1)^{l}k!}{(k-l)!l!}  
\Laplace_{\vec{r}}^{k-l} \big(\hat{u}(\phi)\!\cdot\!\Nabla_{\vec{r}}\big)^{m-2(k-l)-1}\rho_{\nu}(\vec{r},t) 
\end{split}\raisetag{3em}%
\label{eq:Upsilon}%
\end{equation}
with the coefficients 
\begin{equation}
\begin{split}%
A^{(\mu\nu)}_{m,k} &= - \!\int^{\infty}_{0}\!\!\!\!\!\! \dif r\, r^{m+1} U^{(\mu\nu)\prime}_{2}(r) \! 
\int^{2\pi}_{0}\!\!\!\!\!\!\! \dif\phi_{\mathrm{R}}\, g_{\mu\nu}(r,\phi_{\mathrm{R}}) 
\cos(\phi_{\mathrm{R}})^{m-k+1} \sin(\phi_{\mathrm{R}})^{k}  
\end{split}%
\label{eq:A}%
\end{equation}
are gradient expansions of the translational densities $\rho_{\mu}(\vec{r},t)$. 
The four terms on the right-hand side of Eq.\ \eqref{eq:DynamischeGlg} describe entropic translational and rotational diffusion, drift caused by the particles' motility, and diffusive transport resulting from interactions of the particles, respectively. When the particles are passive, the third term vanishes and the function $\Upsilon_{\mu}(\vec{r},\phi,t)$ reduces to the excess chemical potential corresponding to the particles of species $\mu$. 
The dynamic equations \eqref{eq:DynamischeGlg} are applicable even for active systems far from thermodynamic equilibrium and constitute the first main result of this article.

So far, we approximated the pair-correlation functions by assuming that they are invariant with respect to translations and rotations of the coordinate system and that they have no explicit time-dependence, as well as by neglecting the dependence of $g_{\mu\nu}(r,\phi_{\mathrm{R}}-\phi,\phi'\!-\phi)$ on $\phi'\!-\phi$. 
Furthermore, we assumed short-range particle interactions that allow fast convergence of the gradient expansions \eqref{eq:Fres} and \eqref{eq:Upsilon}. 
Most of these approximations are common when dealing with passive particles. The only exception is the negligence of the $\phi'\!-\phi$-dependence of the pair-correlation functions, which depend only on $r$ when the motilities of all particles vanish. With additional assumptions on the pair-correlation functions, one can reduce the number of independent coefficients $A^{(\mu\nu)}_{m,k}$ in Eqs.\ \eqref{eq:Fres} and \eqref{eq:Upsilon}. These possible further approximations of the pair-correlation functions are described in Appendix \ref{Appendix:A}, but not used in this article. 

We proceed by expanding the orientation-dependent one-particle densities $\rho_{\mu}(\vec{r},\phi,t)$ with respect to the orientation $\phi$, neglecting terms of third or higher order in gradients, and performing a quasi-stationary approximation (see Appendix \ref{Appendix:B} for details). This results in the dynamic equations
\begin{equation}%
\dot{\rho}_{\mu} = \sum^{n}_{\nu=1} \partial_{i}\Big( D^{(\mu\nu)}(\{\rho_{\mu}\}) \partial_{i}\rho_{\nu} \Big) 
\label{eq:rhodynD}%
\end{equation}
for the translational densities $\rho_{\mu}(\vec{r},t)$. Here and in the following, $\partial_{i}$ denotes the $i$th element of the del symbol $\Nabla_{\vec{r}}=(\partial_{1},\partial_{2})=(\partial_{x},\partial_{y})$ and Einstein's summation convention is used for Latin but not for Greek indices. Furthermore, the density-dependent diffusion tensor $\boldsymbol{\mathrm{D}}(\{\rho_{\mu}\})$ has the elements
\begin{equation}%
\begin{split}%
D^{(\mu\nu)}(\{\rho_{\mu}\}) &= \bigg(D^{\mu}_{\mathrm{T}}+\frac{v^{2}_{\mu}}{2 D^{\mu}_{\mathrm{R}}}\bigg)\delta_{\mu\nu} 
+ \frac{\rho_{\mu}}{2}\bigg( a^{(\mu\nu)}_{1} 
- a^{(\mu\nu)}_{0}\frac{v^{\mu}_{0} v_{\mu}}{D^{\mu}_{\mathrm{R}}} \bigg) ,
\end{split}%
\label{eq:Drho}%
\end{equation}
the density-dependent swim speeds $v_{\mu}(\{\rho_{\mu}\})$ are given by 
\begin{equation}
v_{\mu}(\{\rho_{\mu}\}) = v^{\mu}_{0}\bigg(1 - \sum^{n}_{\nu=1}  a^{(\mu\nu)}_{0}\rho_{\nu}\bigg) ,
\label{eq:va}%
\end{equation}
and the coefficients $a^{(\mu\nu)}_{0}$ and $a^{(\mu\nu)}_{1}$ are defined as\footnote{For $v^{\mu}_{0}\to 0$ the pair-correlation functions $g_{\mu\nu}(r,\phi)$ become independent of $\phi$ and the coefficients $A^{(\mu\nu)}_{0,0}$ vanish. Therefore, the coefficients $a^{(\mu\nu)}_{0}$ do not diverge when $v^{\mu}_{0}\to 0$.}
{\allowdisplaybreaks
\begin{align}%
\begin{split}%
a^{(\mu\nu)}_{0}=\beta D^{\mu}_{\mathrm{T}}\frac{A^{(\mu\nu)}_{0,0}}{v^{\mu}_{0}} \;, 
\label{eq:a0A}%
\end{split}\\%
\begin{split}%
a^{(\mu\nu)}_{1}=\beta D^{\mu}_{\mathrm{T}}(A^{(\mu\nu)}_{1,0}+A^{(\mu\nu)}_{1,2}) \;. 
\label{eq:a1A}%
\end{split}%
\end{align}}%
The division of $A^{(\mu\nu)}_{0,0}$ by $v^{\mu}_{0}$ in the definition of the coefficients $a^{(\mu\nu)}_{0}$ is motivated by the fact that the effective swim speeds $v_{\mu}(\{\rho_{\mu}\})$ are proportional to $v^{\mu}_{0}$ in mixtures of active and passive particles \cite{StenhammarWMC2015}. 
Equations \eqref{eq:rhodynD} and \eqref{eq:Drho} constitute the second main result of this article.

On the right-hand side of Eq.\ \eqref{eq:Drho}, the term $D^{\mu}_{\mathrm{T}}\delta_{\mu\nu}$ describes the diffusivity of a free passive particle, the term $a^{(\mu\nu)}_{1}\rho_{\mu}/2$ takes interactions with particles of the same or other species into account, and the two remaining contributions describe the effect of the particles' motilities on the collective diffusion.   
Since the diffusion tensor $\boldsymbol{\mathrm{D}}(\{\rho_{\mu}\})$ and the swim speeds $v_{\mu}(\{\rho_{\mu}\})$ can directly be obtained from particle-resolved computer simulations, Eqs.\ \eqref{eq:Drho} and \eqref{eq:va} allow to determine the coefficients $a^{(\mu\nu)}_{0}$ and $a^{(\mu\nu)}_{1}$:
{\allowdisplaybreaks
\begin{align}%
\begin{split}%
a^{(\mu\nu)}_{0}&= -\frac{1}{v^{\mu}_{0}}\pdif{v_{\mu}}{\rho_{\nu}} \;, 
\label{eq:a0Dv}%
\end{split}\\%
\begin{split}%
a^{(\mu\nu)}_{1}&= \frac{2}{\rho_{\mu}} \bigg(D^{(\mu\nu)}-\bigg(D^{\mu}_{\mathrm{T}}+\frac{v^{2}_{\mu}}{2 D^{\mu}_{\mathrm{R}}}\bigg)\delta_{\mu\nu} \bigg) 
-\frac{v_{\mu}}{D^{\mu}_{\mathrm{R}}}\pdif{v_{\mu}}{\rho_{\nu}}  \;. 
\label{eq:a1Dv}%
\end{split}%
\end{align}}%
Especially for one-component systems, this constitutes an alternative to determining the pair-correlation functions in simulations and calculating $a^{(\mu\nu)}_{0}$ and $a^{(\mu\nu)}_{1}$ using Eqs.\ \eqref{eq:A}, \eqref{eq:a0A}, and \eqref{eq:a1A}.

\section{\label{sec:instability}Dynamical instability}
For certain combinations of the free swim speeds $v^{\mu}_{0}$ and the average particle number densities 
$\bar{\rho}_{\mu}$ the mixture becomes unstable, particle clusters form, and the system phase separates \cite{StenhammarWMC2015}.
In the following, we use the dynamic equations \eqref{eq:rhodynD} to derive the spinodal for the onset of this activity-induced dynamical instability by means of a linear stability analysis.

\subsection{Multicomponent mixtures}
To derive the spinodal, we consider a system with slightly perturbed translational densities $\rho_{\mu}(\vec{r},t)=\bar{\rho}_{\mu}+\delta\rho_{\mu}(\vec{r},t)$, where $\delta\rho_{\mu}(\vec{r},t)$ 
are the small perturbations.
Inserting this perturbation ansatz into Eq.\ \eqref{eq:rhodynD} and linearizing the resulting expression in $\delta\rho_{\mu}(\vec{r},t)$ leads to the dynamic equations for the perturbations
\begin{equation}%
\dot{\delta\rho}_{\mu} = \sum^{n}_{\nu=1} \bar{D}_{\mu\nu} \Laplace \delta\rho_{\nu}  
\label{eq:rhodynDlinearized}%
\end{equation}
with the constant diffusion coefficients $\bar{D}_{\mu\nu}=D^{(\mu\nu)}(\{\bar{\rho}_{\mu}\})$ corresponding to the average densities $\bar{\rho}_{1},\dotsc,\bar{\rho}_{n}$. 
Fourier transforming Eq.\ \eqref{eq:rhodynDlinearized} results in 
\begin{equation}%
\dot{\widetilde{\delta\rho}}_{\mu} = -k^{2} \sum^{n}_{\nu=1} \bar{D}_{\mu\nu} \widetilde{\delta\rho}_{\nu} \;,  
\label{eq:rhodynDlinearizedFT}%
\end{equation}
where $\widetilde{\delta\rho}_{\mu}(\vec{k},t)$ denotes the spatial Fourier transform of $\delta\rho_{\mu}(\vec{r},t)$ and $k=\norm{\vec{k}}$ is the wave number corresponding to the wave vector $\vec{k}$. 
If we define the perturbation vector $\vec{\widetilde{\delta\rho}}=(\widetilde{\delta\rho}_{1},\dotsc,\widetilde{\delta\rho}_{n})^{\mathrm{T}}$ and the constant diffusion tensor $\boldsymbol{\bar{\mathrm{D}}}=\boldsymbol{\mathrm{D}}(\{\bar{\rho}_{\mu}\})=(\bar{D}_{\mu\nu})_{\mu,\nu=1,\dotsc,n}$, the solution of Eq.\ \eqref{eq:rhodynDlinearizedFT} can be written as
\begin{equation}%
\vec{\widetilde{\delta\rho}}(\vec{k},t) = \exp\!\big(\! -k^{2}\boldsymbol{\bar{\mathrm{D}}} t \big) \vec{\widetilde{\delta\rho}}(\vec{k},0)
\label{eq:rhodynDlinearizedFTLsg}%
\end{equation}
with the matrix exponential $\exp(\! -k^{2}\boldsymbol{\bar{\mathrm{D}}} t)$.
An $n$-component mixture of colloidal particles with different motilities is therefore stable if the real parts of all eigenvalues $\lambda_{1},\dotsc,\lambda_{n}$ of $\boldsymbol{\bar{\mathrm{D}}}$ are nonnegative, i.e., 
\begin{equation}%
\Real(\lambda_{i})\geqslant 0 \qquad \forall\, i\in\{1,\dotsc,n\}  \;,
\label{eq:stabilityconditionIV}%
\end{equation}
and unstable otherwise (see Appendix \ref{Appendix:C} for details). 
The spinodal of the dynamical instability is the outer surface that encloses the surfaces which are defined by the implicit functions $\Real(\lambda_{i})=0$ with $i\in\{1,\dotsc,n\}$. 
In the next subsection, we apply the stability criterion \eqref{eq:stabilityconditionIV} to binary mixtures.

\subsection{\label{sec:binarymixtures}Binary mixtures}
We now focus on binary mixtures with two species $\mu\in\{\mathrm{A},\mathrm{B}\}$ of particles.  
The constant diffusion tensor $\boldsymbol{\bar{\mathrm{D}}}$ is then a real $2\!\times\!2$-dimensional matrix 
\begin{equation}
\boldsymbol{\bar{\mathrm{D}}} = 
\begin{pmatrix} \bar{D}_{\mathrm{AA}} & \bar{D}_{\mathrm{AB}} \\
\bar{D}_{\mathrm{BA}} & \bar{D}_{\mathrm{BB}} \end{pmatrix}
\label{eq:Dbinaer}%
\end{equation}
and the spinodal is the outer surface that encloses the two nonoverlapping\footnote{Combining Eqs.\ \eqref{eq:SpinodalI} and \eqref{eq:SpinodalII} leads to the unfulfillable condition $\bar{D}^{2}_{\mathrm{AA}}<\bar{D}^{2}_{\mathrm{AA}}$. This shows that the surfaces described by Eqs.\ \eqref{eq:SpinodalI} and \eqref{eq:SpinodalII} have no points in common.} surfaces given by 
\begin{equation}
\bar{D}_{\mathrm{AA}}=-\bar{D}_{\mathrm{BB}} \quad \wedge \quad
\bar{D}_{\mathrm{AA}}^{2} < - \bar{D}_{\mathrm{AB}} \bar{D}_{\mathrm{BA}}
\label{eq:SpinodalI}%
\end{equation}
and 
\begin{equation}
\bar{D}_{\mathrm{AA}}\bar{D}_{\mathrm{BB}} 
=\bar{D}_{\mathrm{AB}} \bar{D}_{\mathrm{BA}} \;,
\label{eq:SpinodalII}%
\end{equation}
respectively. Equations \eqref{eq:SpinodalI} and \eqref{eq:SpinodalII} describe the stability of binary mixtures of colloidal particles with different motilities and constitute the third main result of this article.  

As can be seen from the eigenvalues 
{\allowdisplaybreaks
\begin{align}%
\begin{split}%
\lambda_{1} &= \frac{1}{2}(\bar{D}_{\mathrm{AA}} + \bar{D}_{\mathrm{BB}}) 
- \frac{1}{2}\sqrt{(\bar{D}_{\mathrm{AA}}-\bar{D}_{\mathrm{BB}})^{2} + 4\bar{D}_{\mathrm{AB}}\bar{D}_{\mathrm{BA}}} \;, 
\label{eq:lambdaI}%
\end{split}\\%
\begin{split}%
\lambda_{2} &= \frac{1}{2}(\bar{D}_{\mathrm{AA}} + \bar{D}_{\mathrm{BB}}) 
+ \frac{1}{2}\sqrt{(\bar{D}_{\mathrm{AA}}-\bar{D}_{\mathrm{BB}})^{2} + 4\bar{D}_{\mathrm{AB}}\bar{D}_{\mathrm{BA}}} 
\label{eq:lambdaII}%
\end{split}%
\end{align}}%
and corresponding eigenvectors 
\begin{equation}
\vec{w}_{1} = \begin{pmatrix} \lambda_{1} - \bar{D}_{\mathrm{BB}} \\ \bar{D}_{\mathrm{BA}} \end{pmatrix} , \qquad  
\vec{w}_{2} = \begin{pmatrix} \lambda_{2} - \bar{D}_{\mathrm{BB}} \\ \bar{D}_{\mathrm{BA}} \end{pmatrix} 
\label{eq:Eigenvektoren}%
\end{equation}
of $\boldsymbol{\bar{\mathrm{D}}}$, respectively, Eqs.\ \eqref{eq:SpinodalI} and \eqref{eq:SpinodalII} correspond to instabilities of different types. Under the condition \eqref{eq:SpinodalI}, the eigenvalues $\lambda_{1}$ and $\lambda_{2}$ are imaginary and complex conjugates of each other. The surface described by Eq.\ \eqref{eq:SpinodalI}, where a complex conjugate pair of eigenvalues of $\boldsymbol{\bar{\mathrm{D}}}$ passes through the imaginary axis, is associated with a Hopf bifurcation. This type of bifurcation is possible for a binary mixture, but not for a one-component system of active or passive colloidal particles.
In contrast, under the condition \eqref{eq:SpinodalII}, $\lambda_{1}$ and $\lambda_{2}$ are real and one of these eigenvalues is zero. 
The surface described by Eq.\ \eqref{eq:SpinodalII}, where one of the two real eigenvalues of $\boldsymbol{\bar{\mathrm{D}}}$ passes through zero, is associated with a stationary bifurcation. 
This type of bifurcation is possible also for one-component systems. 
Hence, there is a fundamental difference between the already widely studied motility-induced phase separation of one-component systems of active particles, which we related to a stationary bifurcation, and the nonequilibrium dynamics of active-passive mixtures, which exhibits also a Hopf bifurcation.  

To study the features of these bifurcations, we now insert the plane-wave ansatz
\begin{equation}
\vec{\delta\rho}(\vec{r},t) = \sum^{2}_{i=1} A_{i} e^{\ii (\vec{k}_{i}\!\cdot\!\vec{r} - \omega_{i}t)} \vec{w}_{i}
\label{eq:Wellenansatz}%
\end{equation}
with the perturbation vector $\vec{\delta\rho}=(\delta\rho_{\mathrm{A}},\delta\rho_{\mathrm{B}})^{\mathrm{T}}$, constants $A_{1},A_{2}\in\C$, imaginary unit $\ii$, wave vectors $\vec{k}_{1},\vec{k}_{2}\in\R^{2}$, and angular frequencies $\omega_{1},\omega_{2}\in\C$ into Eq.\ \eqref{eq:rhodynDlinearized}.\footnote{Note that in the plane-wave ansatz \eqref{eq:Wellenansatz} the perturbation vector $\vec{\delta\rho}$ is complex. The perturbations that could be observed in simulations or experiments correspond to the real part of $\vec{\delta\rho}$.} This results in the dispersion relations 
{\allowdisplaybreaks
\begin{align}%
\begin{split}%
\omega_{1} &= -\ii \lambda_{1}k^{2}_{1} \;, 
\label{eq:omegaI}%
\end{split}\\%
\begin{split}%
\omega_{2} &= -\ii \lambda_{2}k^{2}_{2} 
\label{eq:omegaII}%
\end{split}%
\end{align}}%
with $k_{1}=\norm{\vec{k}_{1}}$ and $k_{2}=\norm{\vec{k}_{2}}$. When decomposing the eigenvalues $\lambda_{i}=\Real(\lambda_{i})+\ii\Imag(\lambda_{i})$ with $i\in\{1,2\}$ into their real parts $\Real(\lambda_{i})$ and imaginary parts $\Imag(\lambda_{i})$, where $\Imag(\lambda_{1})=-\Imag(\lambda_{2})$, and inserting the dispersion relations \eqref{eq:omegaI} and \eqref{eq:omegaII} into the ansatz \eqref{eq:Wellenansatz}, we obtain 
\begin{equation}
\vec{\delta\rho}(\vec{r},t) = \sum^{2}_{i=1} A_{i} e^{\ii (\vec{k}_{i}\!\cdot\!\vec{r} - \Imag(\lambda_{i})k^{2}_{i}t)} e^{-\Real(\lambda_{i})k^{2}_{i}t} \vec{w}_{i} \;.
\label{eq:Wellenloesung}%
\end{equation}
From this equation it is obvious that -- in accordance with the stability criterion \eqref{eq:stabilityconditionIV} -- the amplitudes of small spatially harmonic perturbations are exponentially growing with time when at least one of the eigenvalues of $\boldsymbol{\bar{\mathrm{D}}}$ has a negative real part. 
Furthermore, we see that the perturbations are traveling plane waves for $\Imag(\lambda_{1})\neq 0$, i.e., for the Hopf bifurcation associated with Eq.\ \eqref{eq:SpinodalI}, whereas the perturbations are static for $\Imag(\lambda_{1})=0$, as holds for the stationary bifurcation associated with Eq.\ \eqref{eq:SpinodalII}. 
This is an important difference between the two types of bifurcations and helps to distinguish the two underlying instabilities in simulations and experiments. 
The speed of a traveling spatially harmonic perturbation with wave number $k$ is $\Imag(\lambda_{2})k$.
Note that this expression holds only for small $k$, since the continuum description by the field theory \eqref{eq:rhodynD} breaks down for $k\gtrsim 2\pi/(R_{\mathrm{A}}+R_{\mathrm{B}})$, where $R_{\mathrm{A}}$ and $R_{\mathrm{B}}$ are the radii of particles of species A and B, respectively.

\subsubsection{\label{sec:APmixtures}Active-passive mixtures}
In the following, we assume that all particles have the same radius $R$ and thus identical translational diffusion coefficients $D_{\mathrm{T}}=D^{\mathrm{A}}_{\mathrm{T}}=D^{\mathrm{B}}_{\mathrm{T}}$ as well as identical rotational diffusion coefficients $D_{\mathrm{R}}=D^{\mathrm{A}}_{\mathrm{R}}=D^{\mathrm{B}}_{\mathrm{R}}$.
The elements of the diffusion tensor \eqref{eq:Dbinaer} then simplify to 
{\allowdisplaybreaks
\begin{align}%
\begin{split}%
\bar{D}_{\mathrm{AA}} &= D_{\mathrm{T}} + \frac{\bar{v}^{2}_{\mathrm{A}}}{2 D_{\mathrm{R}}} + \frac{\bar{\rho}_{\mathrm{A}}}{2}\bigg( a^{(\mathrm{AA})}_{1} 
- a^{(\mathrm{AA})}_{0}\frac{v^{\mathrm{A}}_{0} \bar{v}_{\mathrm{A}}}{D_{\mathrm{R}}} \bigg) , 
\label{eq:DbinaerABa}%
\end{split}\\%
\begin{split}%
\bar{D}_{\mathrm{AB}} &= \frac{\bar{\rho}_{\mathrm{A}}}{2}\bigg( a^{(\mathrm{AB})}_{1} - a^{(\mathrm{AB})}_{0}\frac{v^{\mathrm{A}}_{0} \bar{v}_{\mathrm{A}}}{D_{\mathrm{R}}} \bigg) ,
\label{eq:DbinaerABb}%
\end{split}\\%
\begin{split}%
\bar{D}_{\mathrm{BA}} &= \frac{\bar{\rho}_{\mathrm{B}}}{2}\bigg( a^{(\mathrm{BA})}_{1} - a^{(\mathrm{BA})}_{0}\frac{v^{\mathrm{B}}_{0} \bar{v}_{\mathrm{B}}}{D_{\mathrm{R}}} \bigg) , 
\label{eq:DbinaerABc}%
\end{split}\\%
\begin{split}%
\bar{D}_{\mathrm{BB}} &= D_{\mathrm{T}} + \frac{\bar{v}^{2}_{\mathrm{B}}}{2 D_{\mathrm{R}}} + \frac{\bar{\rho}_{\mathrm{B}}}{2}\bigg( a^{(\mathrm{BB})}_{1} 
- a^{(\mathrm{BB})}_{0}\frac{v^{\mathrm{B}}_{0} \bar{v}_{\mathrm{B}}}{D_{\mathrm{R}}} \bigg) 
\label{eq:DbinaerABd}%
\end{split}%
\end{align}}%
with the effective swim speeds 
{\allowdisplaybreaks
\begin{align}%
\begin{split}%
\bar{v}_{\mathrm{A}} &= v^{\mathrm{A}}_{0}(1 - a^{(\mathrm{AA})}_{0}\bar{\rho}_{\mathrm{A}} - a^{(\mathrm{AB})}_{0}\bar{\rho}_{\mathrm{B}}) \;, 
\label{eq:vbinaerABa}%
\end{split}\\%
\begin{split}%
\bar{v}_{\mathrm{B}} &= v^{\mathrm{B}}_{0}(1 - a^{(\mathrm{BA})}_{0}\bar{\rho}_{\mathrm{A}} - a^{(\mathrm{BB})}_{0}\bar{\rho}_{\mathrm{B}}) \;.
\label{eq:vbinaerABb}%
\end{split}%
\end{align}}%

In the special case of a mixture of active (A) and passive (P) particles with $\mu\in\{\mathrm{A},\mathrm{P}\}$, Eqs.\ \eqref{eq:DbinaerABa}-\eqref{eq:vbinaerABb} further simplify due to $\bar{v}_{\mathrm{P}}=v^{\mathrm{P}}_{0}=0$.
With the notation $\bar{v}=\bar{v}_{\mathrm{A}}$ and $v_{0}=v^{\mathrm{A}}_{0}$ this results in 
{\allowdisplaybreaks
\begin{align}%
\begin{split}%
\bar{D}_{\mathrm{AA}} &= D_{\mathrm{T}} + \frac{\bar{v}^{2}}{2 D_{\mathrm{R}}} + \frac{\bar{\rho}_{\mathrm{A}}}{2}\bigg( a^{(\mathrm{AA})}_{1} 
- a^{(\mathrm{AA})}_{0}\frac{v_{0} \bar{v}}{D_{\mathrm{R}}} \bigg) , 
\label{eq:DbinaerAPa}%
\end{split}\\%
\begin{split}%
\bar{D}_{\mathrm{AP}} &= \frac{\bar{\rho}_{\mathrm{A}}}{2}\bigg( a^{(\mathrm{AP})}_{1} - a^{(\mathrm{AP})}_{0}\frac{v_{0} \bar{v}}{D_{\mathrm{R}}} \bigg) , 
\label{eq:DbinaerAPb}%
\end{split}\\%
\begin{split}%
\bar{D}_{\mathrm{PA}} &= \frac{a^{(\mathrm{PA})}_{1}}{2}\bar{\rho}_{\mathrm{P}} \,, 
\label{eq:DbinaerAPc}%
\end{split}\\%
\begin{split}%
\bar{D}_{\mathrm{PP}} &= D_{\mathrm{T}} + \frac{a^{(\mathrm{PP})}_{1}}{2}\bar{\rho}_{\mathrm{P}} 
\label{eq:DbinaerAPd}%
\end{split}%
\end{align}}%
and 
\begin{equation}
\bar{v} = v_{0}(1 - a^{(\mathrm{AA})}_{0}\bar{\rho}_{\mathrm{A}} - a^{(\mathrm{AP})}_{0}\bar{\rho}_{\mathrm{P}}) \;.
\label{eq:vbinaerAP}%
\end{equation}
Equations \eqref{eq:DbinaerAPa}-\eqref{eq:DbinaerAPd} show that for purely repulsive interactions, where $a^{(\mathrm{PA})}_{1}>0$ and $a^{(\mathrm{PP})}_{1}>0$, the passive particles are always stable and that the clustering is triggered by a dynamical instability of the active particles. 
According to Eqs.\ \eqref{eq:SpinodalI} and \eqref{eq:SpinodalII}, the spinodal for the instability of the active-passive mixture is defined by
\begin{equation}
\bar{D}_{\mathrm{AA}}=-\bar{D}_{\mathrm{PP}} \quad \wedge \quad
\bar{D}_{\mathrm{AA}}^{2} < - \bar{D}_{\mathrm{AP}} \bar{D}_{\mathrm{PA}}
\label{eq:SpinodalAPI}%
\end{equation}
and 
\begin{equation}
\bar{D}_{\mathrm{AA}}\bar{D}_{\mathrm{PP}} 
=\bar{D}_{\mathrm{AP}} \bar{D}_{\mathrm{PA}} \;.
\label{eq:SpinodalAPII}%
\end{equation}

\subsubsection{Large $v_{0}$}
We now consider large values $v_{0}\to\infty$ of the self-propulsion speed of a free particle $v_{0}$, but not too large particle densities $\bar{\rho}_{\mathrm{A}}$ and $\bar{\rho}_{\mathrm{P}}$ so that $\bar{v}\to\infty$. Assuming that the coefficients $a^{(\mathrm{AA})}_{0}$, $a^{(\mathrm{AP})}_{0}$, $a^{(\mathrm{AA})}_{1}$, $a^{(\mathrm{AP})}_{1}$, $a^{(\mathrm{PA})}_{1}$, and $a^{(\mathrm{PP})}_{1}$ neither vanish nor diverge when $v_{0}\to\infty$, Eq.\ \eqref{eq:SpinodalAPI} then no longer has a solution, whereas Eq.\ \eqref{eq:SpinodalAPII} simplifies to 
\begin{equation}
\bar{\rho}_{\mathrm{A}} = \frac{2D_{\mathrm{T}} - (2D_{\mathrm{T}}a^{(\mathrm{AP})}_{0}-a^{(\mathrm{PP})}_{1})\bar{\rho}_{\mathrm{P}} - a^{(\mathrm{AP})}_{0}a^{(\mathrm{PP})}_{1}\bar{\rho}^{2}_{\mathrm{P}}}{4D_{\mathrm{T}}a^{(\mathrm{AA})}_{0} 
+ (2a^{(\mathrm{AA})}_{0}a^{(\mathrm{PP})}_{1} - a^{(\mathrm{AP})}_{0}a^{(\mathrm{PA})}_{1})\bar{\rho}_{\mathrm{P}}} \;.
\label{eq:LargePecletI}%
\end{equation}
This means that only the stationary bifurcation but not the Hopf bifurcation occurs in the limit $v_{0}\to\infty$. 
If we set $a^{(\mathrm{PA})}_{1}=a^{(\mathrm{PP})}_{1}=0$ in Eq.\ \eqref{eq:LargePecletI}, it further reduces to
\begin{equation}
\bar{\rho}_{\mathrm{A}}=\frac{1-a^{(\mathrm{AP})}_{0}\bar{\rho}_{\mathrm{P}}}{2a^{(\mathrm{AA})}_{0}} \;,
\label{eq:LargePecletII}%
\end{equation}
which is a simple condition for the low-density branch of the spinodal, as found previously in Ref.\ \cite{StenhammarWMC2015}.

\subsection{One-component systems}
Another limiting case of Eqs.\ \eqref{eq:SpinodalAPI} and \eqref{eq:SpinodalAPII} is that of a one-component system of active particles where $\bar{\rho}_{\mathrm{P}}=0$. 
Also in this case, Eq.\ \eqref{eq:SpinodalAPI} has no solution so that the Hopf bifurcation with its moving perturbations cannot occur.
The stationary bifurcation, in contrast, is still possible. For this bifurcation we obtain from Eq.\ \eqref{eq:SpinodalAPII} the spinodal condition  
\begin{equation}
\bar{D}_{\mathrm{AA}} = D_{\mathrm{T}} + \frac{\bar{v}^{2}}{2 D_{\mathrm{R}}} + \frac{\bar{\rho}_{\mathrm{A}}}{2}\bigg( a^{(\mathrm{AA})}_{1} 
- a^{(\mathrm{AA})}_{0}\frac{v_{0} \bar{v}}{D_{\mathrm{R}}} \bigg) = 0
\label{eq:SpinodaleEinkomponentig}%
\end{equation}
with the effective speed of the active particles 
\begin{equation}
\bar{v} = v_{0}(1 - a^{(\mathrm{AA})}_{0}\bar{\rho}_{\mathrm{A}}) \;.
\label{eq:vEinkomponentig}%
\end{equation}
When choosing $a^{(\mathrm{AA})}_{1}=0$, Eq.\ \eqref{eq:SpinodaleEinkomponentig} becomes equivalent to the simpler spinodal condition that has been proposed in Ref.\ \cite{BialkeLS2013}.
This approximation is, however, in general not applicable. The term proportional to $a^{(\mathrm{AA})}_{1}$ in the diffusion coefficient $\bar{D}_{\mathrm{AA}}$ takes the density-dependence of the particles' collective diffusion into account that originates from their interactions and is present also if the particles are passive ($v_{0}=0$).
Further below we show that neglecting this term has a strong influence on the predicted spinodal. 

If all particles interact purely repulsively, the onset of the dynamical instability requires a sufficiently large value of the low-density self-propulsion speed $v_{0}$. This value depends on $\bar{\rho}_{\mathrm{A}}$: For small values of $\bar{\rho}_{\mathrm{A}}$, the instability cannot occur at all. 
Considering $a^{(\mathrm{AA})}_{0}$ and $a^{(\mathrm{AA})}_{1}$ as nonzero finite constants, which is -- as we will see in Sec.\ \ref{sec:Koeffizienten} -- only an approximation, the minimal swim speed $v_{0,\mathrm{min}}$ of a free particle that allows for cluster formation is given by
\begin{equation}
\begin{split}%
v^{2}_{0,\mathrm{min}} &= D_{\mathrm{R}}\frac{2D_{\mathrm{T}} + a^{(\mathrm{AA})}_{1}\bar{\rho}_{\mathrm{A,min}}}{-1 + 3a^{(\mathrm{AA})}_{0}\bar{\rho}_{\mathrm{A,min}} - 2a^{(\mathrm{AA})2}_{0}\bar{\rho}^{2}_{\mathrm{A,min}}} \\
&= 8D_{\mathrm{T}}D_{\mathrm{R}} + 3D_{\mathrm{R}}\frac{a^{(\mathrm{AA})}_{1}}{a^{(\mathrm{AA})}_{0}} 
+4D_{\mathrm{R}}\sqrt{4D^{2}_{\mathrm{T}} +3D_{\mathrm{T}}\frac{a^{(\mathrm{AA})}_{1}}{a^{(\mathrm{AA})}_{0}} + \frac{1}{2}\bigg(\frac{a^{(\mathrm{AA})}_{1}}{a^{(\mathrm{AA})}_{0}}\bigg)^{2}} 
\end{split}\raisetag{6em}%
\label{eq:CriticalPointII}%
\end{equation}
and corresponds to the active-particle density
\begin{equation}
\begin{split}%
\bar{\rho}_{\mathrm{A,min}} &= -2\frac{D_{\mathrm{T}}}{a^{(\mathrm{AA})}_{1}} 
+ \frac{1}{a^{(\mathrm{AA})}_{1}} \sqrt{4D^{2}_{\mathrm{T}} +3D_{\mathrm{T}}\frac{a^{(\mathrm{AA})}_{1}}{a^{(\mathrm{AA})}_{0}} + \frac{1}{2}\bigg(\frac{a^{(\mathrm{AA})}_{1}}{a^{(\mathrm{AA})}_{0}}\bigg)^{2}} \;. 
\end{split}\raisetag{5em}%
\label{eq:CriticalPointI}%
\end{equation}
This result for the minimal value of $v_{0}$ at the spinodal generalizes the previous result $v_{0,\mathrm{min}}=4\sqrt{D_{\mathrm{T}}D_{\mathrm{R}}}$ of Ref.\ \cite{BialkeLS2013}, which is obtained from Eq.\ \eqref{eq:CriticalPointII} when choosing $a^{(\mathrm{AA})}_{1}=0$ and has been shown to lead to too small values for $v_{0,\mathrm{min}}$ compared to particle-resolved simulations \cite{StenhammarTAMC2013,StenhammarMAC2014}. 
Note that, within the approximations mentioned above, $(\bar{\rho}_{\mathrm{A,min}},v_{0,\mathrm{min}})$ is the critical point of the one-component system of active particles. 

Next, we consider the spinodal condition \eqref{eq:SpinodaleEinkomponentig} in the limit $\bar{v}\to\infty$ and assume that $a^{(\mathrm{AA})}_{0}$ and $a^{(\mathrm{AA})}_{1}$ are finite in this limit. 
This results in the well-known condition for the low-density spinodal branch of a suspension of active Brownian particles 
$\bar{\rho}_{\mathrm{A}}=1/(2a^{(\mathrm{AA})}_{0})$ \cite{CatesT2013}.

\section{\label{sec:simulations}Simulations}
To confirm our analytical results from Sec.\ \ref{sec:instability}, we carried out Brownian dynamics computer simulations of binary mixtures of isometric active and passive colloidal particles in two spatial dimensions. 
These particle-resolved simulations were based on solving the Langevin equations \eqref{eq:LangevinI} and \eqref{eq:LangevinII} numerically and made use of the LAMMPS molecular dynamics package \cite{Plimpton1995}. 
All particles in our simulations had the same diameter $\sigma=2R$, diffusion coefficients $D_{\mathrm{T}}$ and $D_{\mathrm{R}}=3D_{\mathrm{T}}/\sigma^{2}$, and repulsive interaction potential $U_{2}(r)$.
For the latter, we used the Weeks-Chandler-Andersen potential, i.e., the truncated and shifted Lennard-Jones potential 
\begin{equation}
U_{2}(r) = 
\begin{cases}%
4\varepsilon \big( \big(\tfrac{\sigma}{r}\big)^{12} - \big(\tfrac{\sigma}{r}\big)^{6} \big) + \varepsilon \;,& \text{if } r<2^{1/6}\sigma \;, \\
0\;,& \text{if } r\geqslant 2^{1/6}\sigma \;. 
\end{cases}%
\end{equation}
Here, $\varepsilon$ determines the interaction strength as well as the Lennard-Jones time scale $\tau_{\mathrm{LJ}}=\sigma^{2}/(\varepsilon\beta D_{\mathrm{T}})$.
For convenience, we chose $\sigma$, $\tau_{\mathrm{LJ}}$, and $\varepsilon$ as units for length, time, and energy, respectively.
The active particles had the low-density self-propulsion speed $v_{0}=\beta D_{\mathrm{T}}F_{\mathrm{A}}=24\sigma/\tau_{\mathrm{LJ}}$ with the active driving force $F_{\mathrm{A}}=24\varepsilon/\sigma$. 
We studied active-passive mixtures with various total average area fractions $\Phi=\Phi_{\mathrm{A}}+\Phi_{\mathrm{P}}$, fractions of particles that are active $\chi_{\mathrm{A}}=\Phi_{\mathrm{A}}/\Phi$, and P\'eclet numbers $\mathrm{Pe}=v_{0}\sigma/D_{\mathrm{T}}$, where $\Phi_{\mathrm{A}}=\bar{\rho}_{\mathrm{A}}\pi\sigma^{2}/4$ and $\Phi_{\mathrm{P}}=\bar{\rho}_{\mathrm{P}}\pi\sigma^{2}/4$ are the average area fractions of the active and passive particles, respectively.
The P\'eclet number describes the ratio between motility and thermal diffusion of a particle. 
To ensure that the effective particle radius does not depend on the propulsion speed of a free active particle $v_{0}$, we varied $\mathrm{Pe}$ by changing $D_{\mathrm{T}}$ for all particles, while we kept $v_{0}$ constant (see Refs.\ \cite{StenhammarMAC2014,StenhammarWMC2016} for details).
In the limiting case $\chi_{\mathrm{A}}=0$, which corresponds to a system of only passive particles, we chose $D_{\mathrm{T}}=\sigma^{2}/\tau_{\mathrm{LJ}}$.

Our simulations were carried out for particles with homogeneously distributed random initial positions and orientations in a quadratic simulation domain with size $150\sigma\times 150\sigma$ and periodic boundary conditions.
The number of particles was $5730$ for $\Phi=0.2$ and $22918$ for $\Phi=0.8$.
Furthermore, the simulations comprised $5\cdot 10^{7}$ time steps of length $\Delta t = 5\cdot 10^{-5}\tau_{\mathrm{LJ}}$, of which we discarded the first $2\cdot 10^{6}$ time steps.
The duration of a simulation was thus $2500\tau_{\mathrm{LJ}}$.
For $\mathrm{Pe}=100$, where the Brownian time $\tau_{\mathrm{B}}=\sigma^{2}/D_{\mathrm{T}}$ is $\tau_{\mathrm{B}}\approx 4.2\tau_{\mathrm{LJ}}$, this equals $600\tau_{\mathrm{B}}$. 
When the particles have diameter $\sigma=\unit[1]{\micro\metre}$ and translational diffusion coefficient $D_{\mathrm{T}}=k_{\mathrm{B}}T/(3\pi\eta\sigma)$ \cite{Einstein1905c}, and are dispersed in water with dynamic viscosity $\eta=\unit[10^{-3}]{\pascal\,\second}$ and temperature $T=\unit[293]{\kelvin}$, the Brownian time scale is $\tau_{\mathrm{B}}\approx\unit[2.3]{\second}$ and the duration of the simulations corresponds to about $\unit[1400]{\second}$.

\subsection{Pair-correlation functions}
\begin{figure*}[ht]
\includegraphics[width=0.72\linewidth]{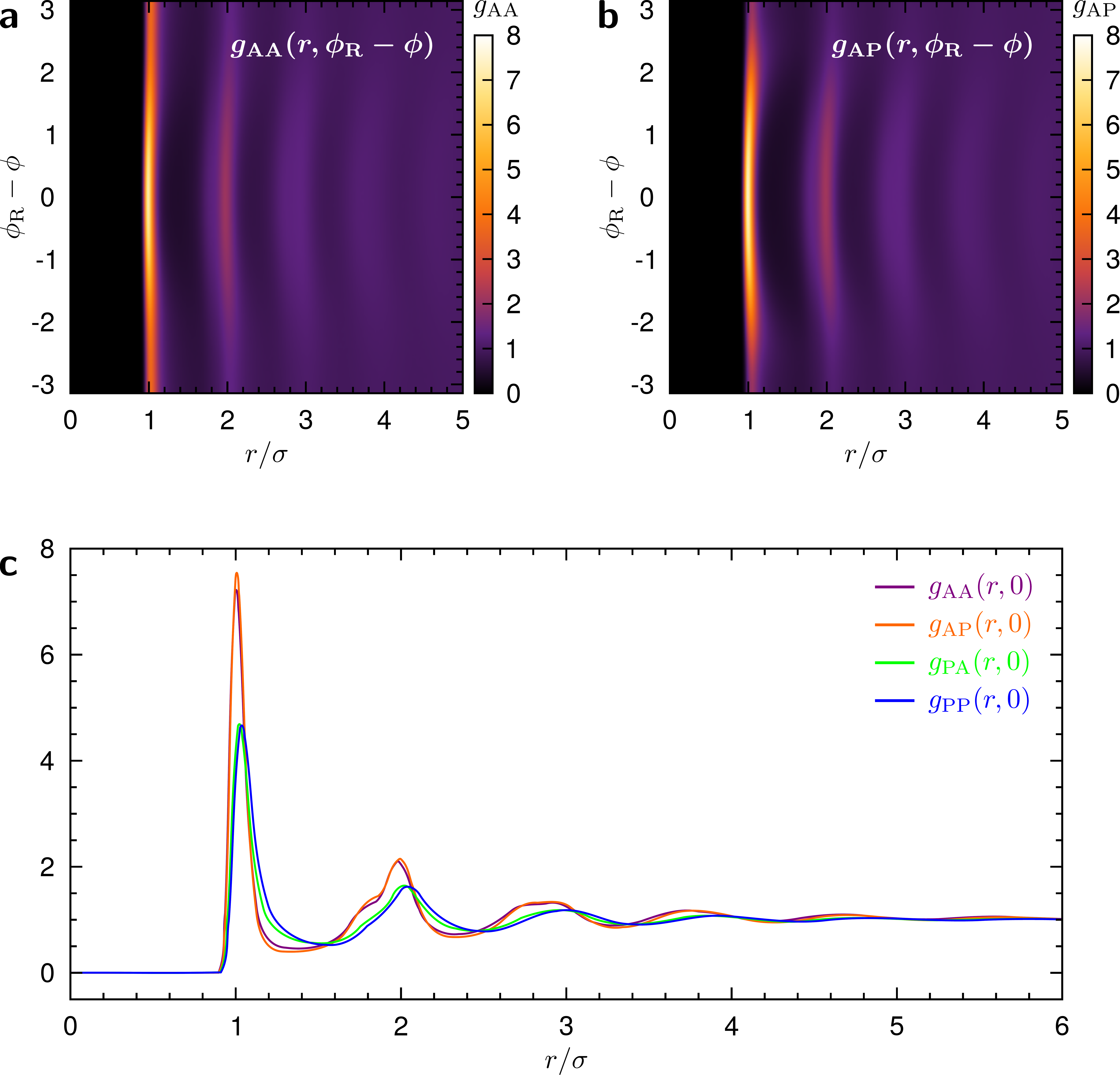}%
\caption{\label{fig:g}Pair-correlation functions (a) $g_{\mathrm{AA}}(r,\phi_{\mathrm{R}}-\phi)$, (b) $g_{\mathrm{AP}}(r,\phi_{\mathrm{R}}-\phi)$, and (c) $g_{\mu\nu}(r,0)$ with $\mu,\nu\in\{\mathrm{A},\mathrm{P}\}$ for P\'{e}clet number $\mathrm{Pe}=50$, total particle area fraction $\Phi=0.6$, and fraction of active particles $\chi_{\mathrm{A}}=0.5$. Note that $g_{\mathrm{PA}}(r,\phi_{\mathrm{R}}-\phi)=g_{\mathrm{PA}}(r,0)$ and $g_{\mathrm{PP}}(r,\phi_{\mathrm{R}}-\phi)=g_{\mathrm{PP}}(r,0)$ for all $\phi_{\mathrm{R}}-\phi$.}%
\end{figure*}

Since the pair-correlation functions $g_{\mathrm{AA}}(r,\phi_{\mathrm{R}}-\phi)$, $g_{\mathrm{AP}}(r,\phi_{\mathrm{R}}-\phi)$, $g_{\mathrm{PA}}(r,\phi_{\mathrm{R}}-\phi)$, and $g_{\mathrm{PP}}(r,\phi_{\mathrm{R}}-\phi)$ are needed to evaluate the coefficients $a^{(\mathrm{AA})}_{0}$, $a^{(\mathrm{AP})}_{0}$, $a^{(\mathrm{AA})}_{1}$, $a^{(\mathrm{AP})}_{1}$, $a^{(\mathrm{PA})}_{1}$, and $a^{(\mathrm{PP})}_{1}$ via Eqs.\ \eqref{eq:A}, \eqref{eq:a0A}, and \eqref{eq:a1A}, we used the simulations described above to determine the pair-correlation functions for $\chi_{\mathrm{A}}=0.5$ and various parameter combinations of $\mathrm{Pe}\in [0,500]$ and $\Phi\in [0,0.8]$.
However, we did not include parameter combinations where the isotropic state of the simulated system was not stable. The resulting pair-correlation functions for $\mathrm{Pe}=50$ and $\Phi=0.6$ are shown in Fig.\ \ref{fig:g}.

For fixed $\phi_{\mathrm{R}}-\phi$, the shapes of these functions of $r$ are qualitatively similar to the typical shape of a radial distribution function for repulsive disks or spheres. However, the maxima and minima of $g_{\mathrm{AA}}(r,\phi_{\mathrm{R}}-\phi)$ and $g_{\mathrm{AP}}(r,\phi_{\mathrm{R}}-\phi)$ as functions of $r$ depend strongly on $\phi_{\mathrm{R}}-\phi$. They are most strongly pronounced for $\phi_{\mathrm{R}}-\phi=0$, corresponding to the active (reference) particle colliding with other particles right in front of it, and become increasingly less pronounced when $\abs{\phi_{\mathrm{R}}-\phi}\leqslant\pi$ increases (see Figs.\ \ref{fig:g}a and \ref{fig:g}b). 
Furthermore, the positions of the maxima and minima move to slightly larger particle separations $r$ for growing $\abs{\phi_{\mathrm{R}}-\phi}\leqslant\pi$.
This is consistent with the bow wave formation in front of a moving active particle reported for one-component systems of active particles in Ref.\ \cite{BialkeLS2013}.
In contrast to $g_{\mathrm{AA}}(r,\phi_{\mathrm{R}}-\phi)$ and $g_{\mathrm{AP}}(r,\phi_{\mathrm{R}}-\phi)$, the pair-correlation functions $g_{\mathrm{PA}}(r,\phi_{\mathrm{R}}-\phi)$ and $g_{\mathrm{PP}}(r,\phi_{\mathrm{R}}-\phi)$ do not depend on $\phi_{\mathrm{R}}-\phi$ so that it is here sufficient to focus on the case $\phi_{\mathrm{R}}-\phi=0$. 

When comparing the four pair-correlation functions for $\phi=\phi_{\mathrm{R}}$, it becomes obvious that $g_{\mathrm{AA}}(r,0)$ and $g_{\mathrm{AP}}(r,0)$, on the one hand, as well as $g_{\mathrm{PA}}(r,0)$ and $g_{\mathrm{PP}}(r,0)$, on the other hand, are very similar.
The main difference between these two pairs of correlation functions is that the maxima of $g_{\mathrm{AA}}(r,0)$ and $g_{\mathrm{AP}}(r,0)$ are significantly larger and shifted to smaller $r$ compared to $g_{\mathrm{PA}}(r,0)$ and $g_{\mathrm{PP}}(r,0)$ (see Fig.\ \ref{fig:g}c).
This is a consequence of the different type of the reference particle, which is active for $g_{\mathrm{AA}}(r,0)$ and $g_{\mathrm{AP}}(r,0)$ but passive for $g_{\mathrm{PA}}(r,0)$ and $g_{\mathrm{PP}}(r,0)$. When the reference particle is active, it pushes other particle ahead of itself so that in front of it (i.e., at $\phi=\phi_{\mathrm{R}}$) on average over time the concentration of particles is larger and the particle separations are smaller than around a passive reference particle.

\subsection{\label{sec:Koeffizienten}Coefficients}
Using Eq.\ \eqref{eq:A} as well as our results for the pair-correlation functions for $\chi_{\mathrm{A}}=0.5$ and various choices of $\mathrm{Pe}$ and $\Phi$, we calculated the corresponding values of the coefficients $a^{(\mathrm{AA})}_{0}$, $a^{(\mathrm{AP})}_{0}$, $a^{(\mathrm{AA})}_{1}$, $a^{(\mathrm{AP})}_{1}$, $a^{(\mathrm{PA})}_{1}$, and $a^{(\mathrm{PP})}_{1}$. 
While for small $\mathrm{Pe}$ we observed a strong dependence of these coefficients on $\mathrm{Pe}$, we found that their dependence on $\mathrm{Pe}$ is negligible for $\mathrm{Pe}>50$. Under the condition $\mathrm{Pe}>50$, the calculated coefficients are approximately given by the expressions
{\allowdisplaybreaks
\begin{align}%
\begin{split}%
a^{(\mathrm{AA})}_{0} &= 0.86 \;, 
\label{eq:aOAAberechnet}%
\end{split}\\%
\begin{split}%
a^{(\mathrm{AP})}_{0} &= 1.04 - 0.67 (\Phi-0.5)^{2} 
\label{eq:aOAPberechnet}%
\end{split}%
\end{align}}%
and
{\allowdisplaybreaks
\begin{align}%
\begin{split}%
a^{(\mathrm{AA})}_{1} &= 24.94 + 11.30 e^{2.62\Phi} \;, 
\label{eq:aIAAberechnet}%
\end{split}\\%
\begin{split}%
a^{(\mathrm{AP})}_{1} &= 26.59 + 4.98 e^{3.50\Phi} \;, 
\label{eq:aIAPberechnet}%
\end{split}\\%
\begin{split}%
a^{(\mathrm{PA})}_{1} &= a^{(\mathrm{AP})}_{1} \,, 
\label{eq:aIPAberechnet}%
\end{split}\\%
\begin{split}%
a^{(\mathrm{PP})}_{1} &= -7.45 + 7.78 e^{3.40\Phi} \;. 
\label{eq:aIPPberechnet}%
\end{split}%
\end{align}}%
Interestingly, the coefficient $a^{(\mathrm{AA})}_{0}$ is independent of the total particle area fraction $\Phi$ and $a^{(\mathrm{AP})}_{0}$ has only a moderate dependence on $\Phi$, whereas the other coefficients depend strongly on $\Phi$. 

With Eqs.\ \eqref{eq:aOAAberechnet} and \eqref{eq:aOAPberechnet} the effective swim speed of an active particle \eqref{eq:vbinaerAP} becomes  
$\bar{v} = v_{0}(1 - 0.86\bar{\rho}_{\mathrm{A}} - (1.04 - 0.67 (\Phi-0.5)^{2})\bar{\rho}_{\mathrm{P}})$ or, equivalently, $\bar{v} = v_{0}(1 - 1.09\Phi_{\mathrm{A}} - (1.32 - 0.85 (\Phi-0.5)^{2})\Phi_{\mathrm{P}})$. This is close\footnote{For $\Phi_{\mathrm{A}},\Phi_{\mathrm{P}}\geqslant 0$ and $\Phi_{\mathrm{A}}+\Phi_{\mathrm{P}}\leqslant\Phi_{\mathrm{cp}}$ with the close-packing area fraction $\Phi_{\mathrm{cp}}=\pi/(2\sqrt{3})$, the maximal difference of both expressions for $\bar{v}$ is less than $0.061v_{0}$.} to the result $\bar{v} = v_{0}(1 - 1.08\Phi_{\mathrm{A}} - 1.21\Phi_{\mathrm{P}})$ from Ref.\ \cite{StenhammarWMC2015}, which has been obtained from particle-resolved simulations for different $\Phi_{\mathrm{A}}$ and $\Phi_{\mathrm{P}}$ by calculating $\bar{v}(\Phi_{\mathrm{A}},\Phi_{\mathrm{P}})$ directly as the time-averaged speed of an active particle. 
For $\Phi_{\mathrm{P}}=0$, our expression reduces to $\bar{v} = v_{0}(1 - 1.09\Phi_{\mathrm{A}})$, which is similar to the corresponding result $\bar{v} = v_{0}(1 - 1.05\Phi_{\mathrm{A}})$ from Refs.\ \cite{StenhammarTAMC2013,StenhammarMAC2014}. 

The equality of $a^{(\mathrm{AP})}_{1}$ and $a^{(\mathrm{PA})}_{1}$ in Eq.\ \eqref{eq:aIPAberechnet} is a consequence of using the same interaction potential for all particles and considering only systems with $\chi_{\mathrm{A}}=0.5$ when calculating the pair-correlation functions. These special conditions lead to $\int^{2\pi}_{0}\!\!\! \dif\phi\, g_{\mathrm{AP}}(r,\phi)=\int^{2\pi}_{0}\!\!\! \dif\phi\, g_{\mathrm{PA}}(r,\phi)$ and thus $a^{(\mathrm{AP})}_{1}=a^{(\mathrm{PA})}_{1}$ here, but they do not hold in general.

\subsection{Dynamical state diagram}
Based on our simulations, we now consider the dynamical state diagram of a binary mixture of isometric active and passive colloidal particles with the same size and the same purely repulsive interactions in two spatial dimensions. For small motilities or low concentrations of active particles, the steady state of such a mixture is homogeneous. 
In contrast, when the P\'eclet number $\mathrm{Pe}$ and the average area fraction of the active particles $\Phi_{\mathrm{A}}$ are sufficiently large, the homogeneous state becomes unstable and particle clusters form \cite{StenhammarWMC2015}. 

\begin{figure*}[ht]
\includegraphics[width=\linewidth]{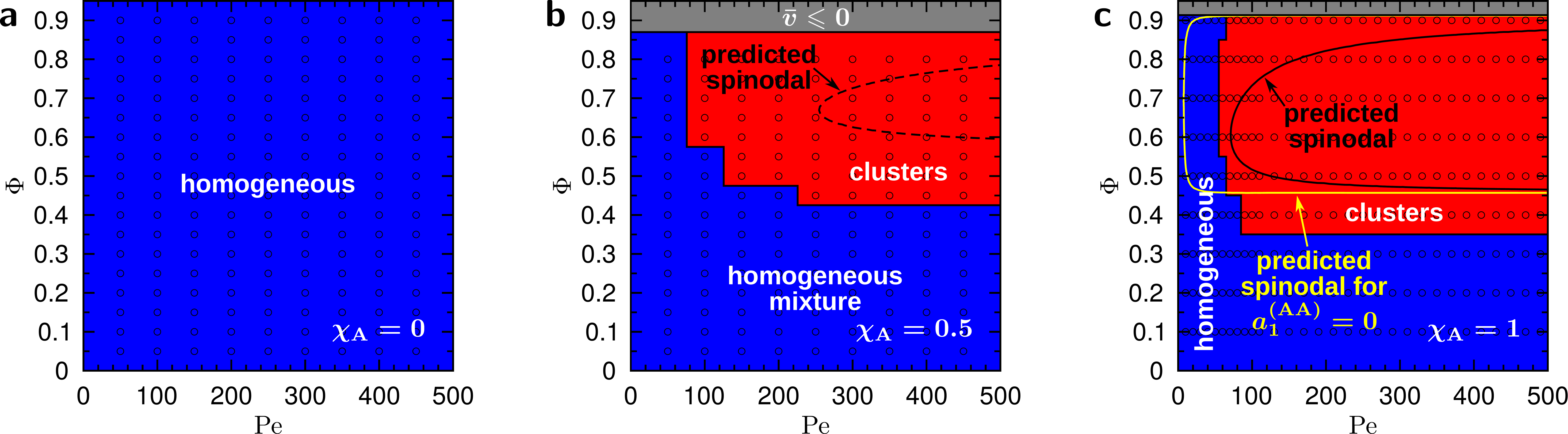}%
\caption{\label{fig:Spinodale}Theoretical predictions (spinodal curves) and simulation results (circles) for the dynamical state diagram of a binary mixture of isometric repulsive active and passive colloidal particles for (a) $\chi_{\mathrm{A}}=0$, (b) $\chi_{\mathrm{A}}=0.5$, and (c) $\chi_{\mathrm{A}}=1$. In the stable region (blue) the system has a homogeneous steady state, whereas in the unstable region (red) particle clusters can be observed. The predicted spinodals for the onset of a dynamical instability in the system correspond to Eq.\ \eqref{eq:SpinodalAPI} (dashed black curve) and Eq.\ \eqref{eq:SpinodalAPII} (solid black curve). In (c), where Eq.\ \eqref{eq:SpinodalAPII} reduces to Eq.\ \eqref{eq:SpinodaleEinkomponentig}, the solid yellow curve corresponds to setting $a^{(\mathrm{AA})}_{1}=0$ in Eq.\ \eqref{eq:SpinodaleEinkomponentig}. Regions with $\bar{v}\leqslant 0$ (gray) agree well with the regions that start at the close-packing area fraction $\Phi_{\mathrm{cp}}=\pi/(2\sqrt{3})\approx 0.907$ and extend to even larger area fractions, where the active particles become effectively trapped.} 
\end{figure*}

Figure \ref{fig:Spinodale} shows the dynamical state diagram of a one-component system of passive particles ($\chi_{\mathrm{A}}=0$), a binary mixture with the same number of active and passive particles ($\chi_{\mathrm{A}}=0.5$), and a one-component system of active particles ($\chi_{\mathrm{A}}=1$) in the $\mathrm{Pe}$-$\Phi$ plane, where the stable and unstable regions were determined by visual inspection.
As expected, the system is homogeneous for all $\mathrm{Pe}$ and $\Phi$ when there are no active particles (see Fig.\ \ref{fig:Spinodale}a), whereas the state diagram of systems with half of the particles or all particles being active contains both stable and unstable regions (see Figs.\ \ref{fig:Spinodale}b and \ref{fig:Spinodale}c).
When $\chi_{\mathrm{A}}>0$, Eq.\ \eqref{eq:vbinaerAP} predicts $\bar{v}\leqslant 0$ for the largest particle packing fractions. The corresponding region in the state diagram is in good agreement with the region where the total particle area fraction $\Phi$ is near the close-packing area fraction $\Phi_{\mathrm{cp}}=\pi/(2\sqrt{3})\approx 0.907$ or even larger so that effective motion of the active particles is not possible. 

In the case of an active-passive mixture with $\chi_{\mathrm{A}}=0.5$, clustering of the particles requires $\mathrm{Pe}>50$ and $\Phi>0.4$ (see Fig.\ \ref{fig:Spinodale}b).
Evidently, the predicted spinodal for the onset of a dynamical instability in the mixture, which is in general given by Eqs.\ \eqref{eq:SpinodalAPI} and \eqref{eq:SpinodalAPII} with the parameters \eqref{eq:aOAAberechnet}-\eqref{eq:aIPPberechnet}, is in accordance with the border between the stable and unstable regions that we observed in the simulations. 
Here, only Eq.\ \eqref{eq:SpinodalAPI} describes the spinodal curve, whereas Eq.\ \eqref{eq:SpinodalAPII} has no solution in the parameter range of the state diagram. 
The appearance of the predicted spinodal clearly inside the actual one can partly be explained by the fact that -- due to the high noise level in the simulations -- in large parts of the simulated instability region clustering occurs through nucleation and growth instead of spinodal decomposition. 

The state diagram of one-component systems of active particles ($\chi_{\mathrm{A}}=1$) is qualitatively similar to that of active-passive mixtures with $\chi_{\mathrm{A}}=0.5$ (see Fig.\ \ref{fig:Spinodale}c). Also in the case of solely active particles, the occurrence of a motility-induced instability requires $\mathrm{Pe}>50$, but in this case an instability can already be observed for $\Phi$ down to $\Phi\approx 0.4$ \cite{StenhammarMAC2014}. For $\chi_{\mathrm{A}}=1$, Eq.\ \eqref{eq:SpinodalAPI} has no solution and the spinodal curve is entirely described by Eq.\ \eqref{eq:SpinodalAPII} or, equivalently, Eq.\ \eqref{eq:SpinodaleEinkomponentig}. The predicted spinodal is in rather good agreement with that directly obtained from the simulations. Especially the minimal values of $\mathrm{Pe}$ required for the instability to occur, that are associated with the critical points of the spinodals, are very close together. 
When $a^{(\mathrm{AA})}_{1}=0$ is chosen in Eq.\ \eqref{eq:SpinodaleEinkomponentig}, the predicted spinodal includes too small values of $\mathrm{Pe}$ down to $4\sqrt{3}\approx 6.9$ and the agreement with the simulation results is significantly worse. This shows that the density-dependence of the particles' collective diffusion has a strong influence on the predicted spinodal and should not be neglected. 
Setting $a^{(\mathrm{AA})}_{1}=a^{(\mathrm{AP})}_{1}=a^{(\mathrm{PA})}_{1}=a^{(\mathrm{PP})}_{1}=0$ in the case of a mixture with $\chi_{\mathrm{A}}=0.5$ leads to an even stronger deterioration of the predictions. Then the predicted spinodal not only includes regions of too small $\mathrm{Pe}$, but also corresponds to the wrong type of instability.  

Remarkably, through Eqs.\ \eqref{eq:SpinodalAPI} and \eqref{eq:SpinodalAPII}, which describe the spinodals for the active-passive mixtures with $\chi_{\mathrm{A}}=0.5$ and for the purely active systems with $\chi_{\mathrm{A}}=1$, respectively, also these spinodals are associated with different types of instabilities, including moving perturbations in the first case and static perturbations in the second case. This is in line with the collective dynamics of mixtures of active and passive particles being more violent, with traveling interfaces and strongly dynamic clusters, than that of purely active systems, where clusters only move through particle diffusion \cite{StenhammarWMC2015,WysockiWG2016}. 
Given the different nature of the instabilities for $\chi_{\mathrm{A}}=0.5$ and $\chi_{\mathrm{A}}=1$, and the fact that the mixtures exhibit significantly larger fluctuations, also the ease with which nucleation occurs in the metastable region is likely to be different for one- and two-component systems. This could, at least partly, explain, why the agreement of the predicted spinodals with the simulation results is better for $\chi_{\mathrm{A}}=1$ than for $\chi_{\mathrm{A}}=0.5$.

\section{\label{sec:conclusions}Conclusions}
We studied multicomponent mixtures of interacting colloidal particles with different but constant motilities and isometric shapes that are suspended in an atomic or molecular solvent. 
For this purpose, we derived a mesoscopic field theory that describes the collective nonequilibrium dynamics of such particles in two spatial dimensions. 
By carrying out a stability analysis for this field theory, we obtained equations that describe the onset of a motility-induced dynamical instability in such mixtures. This instability occurs for sufficiently strong self-propulsion and a sufficiently large concentration of the active particles and leads to cluster formation.

After a general treatment of multicomponent mixtures we focused on the important special case of a binary mixture of interacting active and passive colloidal particles.
We found that the instability leading to cluster formation in such an active-passive mixture can be fundamentally different from that occurring in a one-component system of active particles. 
While in the latter type of systems we found only an instability that is associated with a stationary bifurcation and a steady state of static clusters, in an active-passive mixture an instability that is associated with a Hopf bifurcation and moving clusters can also occur. This finding is in excellent agreement with the different dynamics of clusters in both systems that have already been reported before and explains why these systems can behave in very different ways. 
To complement our analytical calculations, we carried out Brownian dynamics computer simulations of mixtures of active and passive particles. The results of these simulations are well in line with our analytical predictions for the spinodal curves. 

An interesting feature of our theory is that it is applicable for both attractive and repulsive particle interactions and even if the interactions are different for particles of different species.
Our theory therefore constitutes a general framework that can be applied to a particular system by choosing the number of particle species, interaction potentials, motilities, etc.\ appropriately.   
In the future it would be useful to extend this theory towards mixtures of particles with anisometric shapes and thus to make its scope of application even wider.

\begin{acknowledgments}
We thank Edgar Knobloch, Hartmut L\"owen, and Uwe Thiele for helpful discussions. 
R.W.\ is funded by the Deutsche Forschungsgemeinschaft (DFG, German Research Foundation) -- WI 4170/3-1. 
J.S.\ acknowledges funding by a Project Grant from the Swedish Research Council (2015-05449).
M.E.C.\ holds a Royal Society Research Professorship.
\end{acknowledgments}

\appendix
\section{\label{Appendix:A}Possible further approximations of the pair-correlation functions}
The number of independent coefficients $A^{(\mu\nu)}_{m,k}$ in Eqs.\ \eqref{eq:Fres} and \eqref{eq:Upsilon} can be reduced by taking into account the orientational dependence of the pair-correlation functions $g_{\mu\nu}(r,\phi)$ only up to second order, i.e., by approximating 
\begin{equation}
\begin{split}%
g_{\mu\nu}(r,\phi) &\approx g^{(\mu\nu)}_{0}(r) +g^{(\mu\nu)}_{1}(r)\cos(\phi) 
+g^{(\mu\nu)}_{2}(r)\cos(2\phi) \;,
\end{split}%
\end{equation}
and by considering rather hard particles with very short-range interactions where the interaction lengths are much smaller than the particle radii $R_{\mu}$. 
With these approximations, the expressions for the coefficients \eqref{eq:A} simplify to 
\begin{equation}
\begin{split}%
A^{(\mu\nu)}_{m,k} \approx  (R_{\mu}+R_{\nu})^{m+1} \bigg(\! & \alpha^{(\mu\nu)}_{0} \! 
\!\int^{2\pi}_{0}\!\!\!\!\!\!\! \dif\phi\, \cos(\phi)^{m-k+1} \sin(\phi)^{k} 
+\alpha^{(\mu\nu)}_{1} \! 
\!\int^{2\pi}_{0}\!\!\!\!\!\!\! \dif\phi\, \cos(\phi)^{m-k+2} \sin(\phi)^{k} \\
&+\alpha^{(\mu\nu)}_{2} \! 
\!\int^{2\pi}_{0}\!\!\!\!\!\!\! \dif\phi\, \cos(2\phi)\cos(\phi)^{m-k+1} \sin(\phi)^{k} \!\bigg)
\end{split}\raisetag{4.5em}%
\label{eq:A_einfach}%
\end{equation}
with 
\begin{equation}
\alpha^{(\mu\nu)}_{i} = -\!\int^{\infty}_{0}\!\!\!\!\!\! \dif r\, U^{(\mu\nu)\prime}_{2}(r) g^{(\mu\nu)}_{i}(r) 
\;, \quad i\in\{0,1,2\} \;.
\label{eq:a_AP}%
\end{equation}
When the particles are hard disks or spheres in a plane, the coefficients $\alpha^{(\mu\nu)}_{i}$ can be expressed in a simpler form. Following a procedure that is a standard trick when deriving an expression for the pressure of a passive hard-sphere fluid \cite{HansenMD2009}, we write $g^{(\mu\nu)}_{i}(r)$ in Eq.\ \eqref{eq:a_AP} as $\exp(-\beta U^{(\mu\nu)}_{2}(r))\exp(\beta U^{(\mu\nu)}_{2}(r))g^{(\mu\nu)}_{i}(r)$ and afterwards use $U^{(\mu\nu)\prime}_{2}(r)\exp(-\beta U^{(\mu\nu)}_{2}(r))=-\beta^{-1}\dif\exp(-\beta U^{(\mu\nu)}_{2}(r))/\dif r$. For hard-sphere interactions, the function $\exp(-\beta U^{(\mu\nu)}_{2}(r))$ becomes the Heaviside step function $\mathrm{H}(r-R_{\mu}-R_{\nu})$. Taking into account that $\dif\mathrm{H}(r)/\dif r=\delta(r)$, the coefficients $\alpha^{(\mu\nu)}_{i}$ can thus be written as
\begin{equation}
\alpha^{(\mu\nu)}_{i} = \frac{1}{\beta} g^{(\mu\nu)}_{i}(R_{\mu}+R_{\nu})^{+} 
\label{eq:a_APhs}%
\end{equation}
with $g^{(\mu\nu)}_{i}(R_{\mu}+R_{\nu})^{+}=\lim_{\epsilon\to 0^{+}}g^{(\mu\nu)}_{i}(R_{\mu}+R_{\nu}+\epsilon)$ for hard particles.
What now remains to determine are the values of $g^{(\mu\nu)}_{i}(R_{\mu}+R_{\nu})^{+}$. They can still depend on the speeds $v^{\mu}_{0}$ and the average particle number densities $\bar{\rho}_{\mu}$, but are not functions of $r$ or $\phi$. 
Furthermore, due to their close relationship to the pair-correlation functions $g_{\mu\nu}(r,\phi)$ it is possible to estimate at least lower and upper bounds for the values of $g^{(\mu\nu)}_{i}(R_{\mu}+R_{\nu})^{+}$. It should also be possible to make good estimates for their dependence on $v^{\mu}_{0}$ and $\bar{\rho}_{\mu}$.

\section{\label{Appendix:B}Details on the derivation of Eq.\ (\ref{eq:rhodynD})}
Following the common notation in Ginzburg-Landau theories for liquid crystals \cite{deGennesP1995,WittkowskiLB2010,WittkowskiLB2011,WittkowskiLB2011b}, it is useful to approximate the orientation-dependent one-particle densities 
\begin{equation}
\begin{split}%
\rho_{\mu}(\vec{r},\phi,t)&\approx \tfrac{1}{2\pi}\rho_{\mu}(\vec{r},t) +\vec{\PP}_{\mu}(\vec{r},t)\!\cdot\!\hat{u}(\phi) 
+\boldsymbol{\QQ}_{\mu}(\vec{r},t)\!:\!\big(\hat{u}(\phi)\!\otimes\!\hat{u}(\phi)\big) 
\end{split}%
\label{eq:rho_expansion}%
\end{equation}
in the dynamic equations \eqref{eq:DynamischeGlg} by the order-parameter fields
{\allowdisplaybreaks
\begin{gather}%
\begin{split}%
\rho_{\mu}(\vec{r},t) &= \!\int^{2\pi}_{0}\!\!\!\!\!\!\! \dif\phi\, \rho_{\mu}(\vec{r},\phi,t) \;, 
\label{eq:rho}%
\end{split}\\%
\begin{split}%
\vec{\PP}_{\mu}(\vec{r},t) &= \frac{1}{\pi}\!\int^{2\pi}_{0}\!\!\!\!\!\!\! \dif\phi\, \rho_{\mu}(\vec{r},\phi,t) \hat{u}(\phi) \;, 
\label{eq:P}%
\end{split}\\%
\begin{split}%
\boldsymbol{\QQ}_{\mu}(\vec{r},t) &= \frac{2}{\pi}\!\int^{2\pi}_{0}\!\!\!\!\!\!\! \dif\phi\, 
\rho_{\mu}(\vec{r},\phi,t) \big(\hat{u}(\phi)\!\otimes\!\hat{u}(\phi)-\tfrac{1}{2}\Eins\big) \,. 
\label{eq:Q}%
\end{split}%
\end{gather}}%
These order-parameter fields are the translational densities $\rho_{\mu}(\vec{r},t)$, the local polarizations $\vec{\PP}_{\mu}(\vec{r},t)$, and the symmetric and traceless nematic tensors $\boldsymbol{\QQ}_{\mu}(\vec{r},t)$. 
The corresponding \textit{dynamic equations for the order-parameter fields} are\footnote{In the following, the Einstein summation convention is used for Latin indices. There is, however, no implicit summation over Greek indices.} 
\begin{align}
\dot{\rho}_{\mu} + \partial_{i}J^{\rho_{\mu}}_{i} &= 0 \;, \label{eq:rho_dynamic} \\
\dot{\PP}_{\mu,i} + \Phi^{\PP_{\mu}}_{i} &= 0 \;, \label{eq:P_dynamic} \\
\dot{\QQ}_{\mu,ij} + \Phi^{\QQ_{\mu}}_{ij} &= 0 \label{eq:Q_dynamic}
\end{align}
with the currents $\vec{J}^{\rho_{\mu}}(\vec{r},t)$ and the quasi-currents\footnote{Notice that the translational densities $\rho_{\mu}(\vec{r},t)$ are conserved quantities, whereas the local polarizations $\vec{\PP}_{\mu}(\vec{r},t)$ and the nematic tensors $\boldsymbol{\QQ}_{\mu}(\vec{r},t)$ are not conserved.} $\vec{\Phi}^{\PP_{\mu}}(\vec{r},t)$ and $\boldsymbol{\Phi}^{\QQ_{\mu}}(\vec{r},t)$.  
Up to second order in gradients, these currents and quasi-currents are given by 
{\allowdisplaybreaks
\begin{align}%
\begin{split}%
J^{\rho_{\mu}}_{i}&= \pi v_{\mu} \PP_{\mu,i} - D^{\mu}_{\mathrm{T}} (\partial_{i}\rho_{\mu}) 
-\frac{\beta D^{\mu}_{\mathrm{T}}}{2} \sum^{n}_{\nu=1} \Big( (A^{(\mu\nu)}_{1,0}+A^{(\mu\nu)}_{1,2})\rho_{\mu}(\partial_{i}\rho_{\nu}) +\pi(A^{(\mu\nu)}_{1,0}-A^{(\mu\nu)}_{1,2}) \QQ_{\mu,ij}(\partial_{j}\rho_{\nu}) \Big) \,, 
\label{eq:rhodyn}%
\end{split}\\%
\begin{split}%
\Phi^{\PP_{\mu}}_{i}&= \frac{1}{2\pi}\partial_{i}(v_{\mu} \rho_{\mu}) + \frac{1}{2} \partial_{j}(v_{\mu} \QQ_{\mu,ij}) 
+D^{\mu}_{\mathrm{R}} \PP_{\mu,i} - D^{\mu}_{\mathrm{T}} \Laplace_{\vec{r}}  \PP_{\mu,i} 
-\frac{\beta D^{\mu}_{\mathrm{T}}}{4} \sum^{n}_{\nu=1} \Big((A^{(\mu\nu)}_{1,0}+3A^{(\mu\nu)}_{1,2})
\partial_{j}\big(\PP_{\mu,i}(\partial_{j}\rho_{\nu})\big) \\
&\quad\:\! + (A^{(\mu\nu)}_{1,0}-A^{(\mu\nu)}_{1,2}) \big(\partial_{i}\big(\PP_{\mu,j}(\partial_{j}\rho_{\nu})\big) 
+ \partial_{j}\big(\PP_{\mu,j}(\partial_{i}\rho_{\nu})\big) \big) \Big) \,, 
\label{eq:Pdyn}%
\end{split}\\%
\begin{split}%
\Phi^{\QQ_{\mu}}_{ij}&= \frac{1}{2}\chi_{ijkl}\partial_{k}(v_{\mu} \PP_{\mu,l}) 
+ 4D^{\mu}_{\mathrm{R}} \QQ_{\mu,ij} -D^{\mu}_{\mathrm{T}} \Laplace_{\vec{r}} \QQ_{\mu,ij} 
-\frac{\beta D^{\mu}_{\mathrm{T}}}{4\pi} \sum^{n}_{\nu=1} \Big((A^{(\mu\nu)}_{1,0}-A^{(\mu\nu)}_{1,2}) 
\chi_{ijkl}\partial_{k}\big(\rho_{\mu}(\partial_{l}\rho_{\nu})\big) \\
&\quad\:\! + 2\pi(A^{(\mu\nu)}_{1,0}+A^{(\mu\nu)}_{1,2})\partial_{k}\big(\QQ_{\mu,ij}(\partial_{k}\rho_{\nu})\big)\Big) 
\label{eq:Qdyn}%
\end{split}%
\end{align}}%
with the density-dependent swim speeds 
\begin{equation}
v_{\mu}(\{\rho_{\mu}\}) = v^{\mu}_{0} - \beta D^{\mu}_{\mathrm{T}} \sum^{n}_{\nu=1}  A^{(\mu\nu)}_{0,0}\rho_{\nu} \;,
\label{eq:swimspeeds}%
\end{equation}
the tensor $\chi_{ijkl}=\delta_{ik}\delta_{jl}+\delta_{il}\delta_{jk}-\delta_{ij}\delta_{kl}$, and $\partial_{i}$ being the $i$th element of the del symbol $\Nabla_{\vec{r}}=(\partial_{1},\partial_{2})=(\partial_{x},\partial_{y})$. 
When considering the limiting case of a one-component system ($n=1$) and setting $A^{(\mu\nu)}_{1,0}=A^{(\mu\nu)}_{1,2}=0$, which means neglecting all effects of the particle interactions except for the slow-down of active particles due to collisions that is taken into account through the density-dependent effective swim speeds \eqref{eq:swimspeeds}, our Eqs.\ \eqref{eq:rho_dynamic}-\eqref{eq:Qdyn} reduce to the corresponding dynamic equations for active Brownian particles in two spatial dimensions of Ref.\ \cite{CatesT2013}.\footnote{Setting $n=1$ and $A^{(\mu\nu)}_{1,0}=A^{(\mu\nu)}_{1,2}=0$ in our Eqs.\ \eqref{eq:rho_dynamic}-\eqref{eq:Qdyn} results in dynamic equations that are equivalent to those that are obtained by setting $\varphi=\rho/(2\pi)$, $d=2$, $\alpha=0$, $D_{t}=\text{const.}$, $D_{r}=\text{const.}$, and $\chi_{abc}=0$ in Eqs.\ (4)-(6) of Ref.\ \cite{CatesT2013}.}  

Since only the translational densities $\rho_{\mu}(\vec{r},t)$ are conserved quantities, their relaxation times are much larger than the relaxation times of $\vec{\PP}_{\mu}(\vec{r},t)$ and $\boldsymbol{\QQ}_{\mu}(\vec{r},t)$.
When the nematic tensors $\boldsymbol{\QQ}_{\mu}(\vec{r},t)$ are the fastest-relaxing order-parameter fields, we can describe the system on a larger time scale where $\dot{\QQ}_{\mu,ij}=0$. 
Equation \eqref{eq:Qdyn} can then be used to express $\boldsymbol{\QQ}_{\mu}(\vec{r},t)$ in terms of gradients of the order-parameter fields. Inserting this expression recursively into Eqs.\ \eqref{eq:rhodyn}-\eqref{eq:Qdyn} and still neglecting all terms of third or higher order in gradients results in the quasi-stationary approximation  
{\allowdisplaybreaks
\begin{align}%
\begin{split}%
J^{\rho_{\mu}}_{i}&= \pi v_{\mu} \PP_{\mu,i} - D^{\mu}_{\mathrm{T}} (\partial_{i}\rho_{\mu}) 
-\frac{\beta D^{\mu}_{\mathrm{T}}}{2}\rho_{\mu} \sum^{n}_{\nu=1} (A^{(\mu\nu)}_{1,0}+A^{(\mu\nu)}_{1,2})(\partial_{i}\rho_{\nu}) \,, 
\label{eq:rhodynQS}%
\end{split}\\%
\begin{split}%
\Phi^{\PP_{\mu}}_{i}&= \frac{1}{2\pi}\partial_{i}(v_{\mu} \rho_{\mu}) 
-\frac{1}{16D^{\mu}_{\mathrm{R}}} \chi_{ijkl}\partial_{j}\big(v_{\mu} \partial_{k}(v_{\mu} \PP_{\mu,l})\big) 
+D^{\mu}_{\mathrm{R}} \PP_{\mu,i} - D^{\mu}_{\mathrm{T}} \Laplace_{\vec{r}}  \PP_{\mu,i}  \\
&\quad\:\! -\frac{\beta D^{\mu}_{\mathrm{T}}}{4} \sum^{n}_{\nu=1} \Big((A^{(\mu\nu)}_{1,0}+3A^{(\mu\nu)}_{1,2})
\partial_{j}\big(\PP_{\mu,i}(\partial_{j}\rho_{\nu})\big)+ (A^{(\mu\nu)}_{1,0}-A^{(\mu\nu)}_{1,2}) \big(\partial_{i}\big(\PP_{\mu,j}(\partial_{j}\rho_{\nu})\big) 
+ \partial_{j}\big(\PP_{\mu,j}(\partial_{i}\rho_{\nu})\big) \big) \Big) \,, 
\label{eq:PdynQS}%
\end{split}\\%
\begin{split}%
\QQ_{\mu,ij} &= -\frac{1}{8 D^{\mu}_{\mathrm{R}}}\chi_{ijkl}\partial_{k}(v_{\mu} \PP_{\mu,l}) 
+\frac{\beta D^{\mu}_{\mathrm{T}}}{16\pi D^{\mu}_{\mathrm{R}}} \sum^{n}_{\nu=1} (A^{(\mu\nu)}_{1,0}-A^{(\mu\nu)}_{1,2}) 
\chi_{ijkl}\partial_{k}\big(\rho_{\mu}(\partial_{l}\rho_{\nu})\big) \,,  
\label{eq:QdynQS}%
\end{split}%
\end{align}}%
which involves dynamic equations for $\rho_{\mu}(\vec{r},t)$ and $\vec{\PP}_{\mu}(\vec{r},t)$ and a constitutive equation that expresses $\boldsymbol{\QQ}_{\mu}(\vec{r},t)$ in terms of $\rho_{\mu}(\vec{r},t)$ and $\vec{\PP}_{\mu}(\vec{r},t)$. 
On the even larger time scale of the relaxation times of the translational densities $\rho_{\mu}(\vec{r},t)$, we can use 
$\dot{\PP}_{\mu,i}=0$ to further simplify our dynamic equations \eqref{eq:rhodynQS}-\eqref{eq:QdynQS}. 
Proceeding with $\vec{\PP}_{\mu}(\vec{r},t)$ in a similar way as with $\boldsymbol{\QQ}_{\mu}(\vec{r},t)$ leads to the full \textit{quasi-stationary approximation} 
{\allowdisplaybreaks
\begin{align}%
\begin{split}%
J^{\rho_{\mu}}_{i}&= -\frac{v_{\mu}}{2 D^{\mu}_{\mathrm{R}}} \partial_{i}(v_{\mu} \rho_{\mu}) 
-D^{\mu}_{\mathrm{T}} (\partial_{i}\rho_{\mu}) 
-\frac{\beta D^{\mu}_{\mathrm{T}}}{2}\rho_{\mu} \sum^{n}_{\nu=1} (A^{(\mu\nu)}_{1,0}+A^{(\mu\nu)}_{1,2})(\partial_{i}\rho_{\nu}) \,, 
\label{eq:rhodynQSS}%
\end{split}\\%
\begin{split}%
\PP_{\mu,i} &= -\frac{1}{2\pi D^{\mu}_{\mathrm{R}}} \partial_{i}(v_{\mu} \rho_{\mu}) \,, 
\label{eq:PdynQSS}%
\end{split}\\%
\begin{split}%
\QQ_{\mu,ij} &= \frac{1}{16\pi D^{\mu 2}_{\mathrm{R}}}\chi_{ijkl}\partial_{k}\big(v_{\mu} \partial_{l}(v_{\mu} \rho_{\mu})\big) 
+\frac{\beta D^{\mu}_{\mathrm{T}}}{16\pi D^{\mu}_{\mathrm{R}}} \sum^{n}_{\nu=1} (A^{(\mu\nu)}_{1,0}-A^{(\mu\nu)}_{1,2}) 
\chi_{ijkl}\partial_{k}\big(\rho_{\mu}(\partial_{l}\rho_{\nu})\big) \,,   
\label{eq:QdynQSS}\raisetag{4em}%
\end{split}%
\end{align}}%
which involves only a dynamic equation for $\rho_{\mu}(\vec{r},t)$ and constitutive equations for the other order-parameter fields. 

The first term on the right-hand side of Eq.\ \eqref{eq:rhodynQSS} is proportional to the local polarizations \eqref{eq:PdynQSS}. This shows that gradients of $v_{\mu}(\{\rho_{\mu}\})\rho_{\mu}(\vec{r},t)$ lead to local polarizations that drive currents of the translational densities. The other two terms on the right-hand side of Eq.\ \eqref{eq:rhodynQSS} describe diffusive currents, where the first one represents the translational diffusion of noninteracting particles and the second one takes interactions with particles of the same and other species into account. As can be seen from the first term on the right-hand side of Eq.\ \eqref{eq:QdynQSS}, gradients of $v_{\mu}(\{\rho_{\mu}\})\rho_{\mu}(\vec{r},t)$ also cause local nematic order. Like the local polarizations described by Eq.\ \eqref{eq:PdynQSS}, the first term on the right-hand side of Eq.\ \eqref{eq:QdynQSS} vanishes for passive particles. In contrast, the second term on the right-hand side of Eq.\ \eqref{eq:QdynQSS} does not in general vanish for passive particles. 
When we set $n=1$ and $A^{(\mu\nu)}_{1,0}=A^{(\mu\nu)}_{1,2}=0$, our Eqs.\ \eqref{eq:rhodynQSS}-\eqref{eq:QdynQSS} corresponds to Eqs.\ (8) and (9) and the unnumbered equation before Eq.\ (8) in Ref.\ \cite{CatesT2013}. Furthermore, in this limit our Eq.\ \eqref{eq:rhodynQSS} reduces to the simpler diffusion equation that is given by Eq.\ (20) in Ref.\ \cite{BialkeLS2013}.

\section{\label{Appendix:C}Stability criterion for mixtures}
From Eq.\ \eqref{eq:rhodynDlinearizedFTLsg} it follows that an $n$-component mixture of colloidal particles with different motilities is stable, if 
\begin{equation}%
\lim_{t\to\infty} \big\lVert\exp\!\big(\! -\boldsymbol{\bar{\mathrm{D}}} t \big)\big\rVert_{\mathrm{F}} < \infty \;,
\label{eq:stabilitycondition}%
\end{equation}
where $\norm{\,\cdot\,}_{\mathrm{F}}$ is the Frobenius norm, and unstable otherwise. 
The constant diffusion tensor $\boldsymbol{\bar{\mathrm{D}}}=\boldsymbol{\bar{\mathrm{D}}}_{0}
+\boldsymbol{\bar{\mathrm{D}}}_{1}$ can be uniquely decomposed into a diagonalizable matrix $\boldsymbol{\bar{\mathrm{D}}}_{0}$ with the same eigenvalues as $\boldsymbol{\bar{\mathrm{D}}}$ and a nilpotent matrix $\boldsymbol{\bar{\mathrm{D}}}_{1}$ that commutes with $\boldsymbol{\bar{\mathrm{D}}}_{0}$ (Jordan-Chevalley decomposition).
Equation \eqref{eq:stabilitycondition} can therefore be written as 
\begin{equation}%
\lim_{t\to\infty} \big\lVert\exp\!\big(\! -\boldsymbol{\bar{\mathrm{D}}}_{0} t \big)
\exp\!\big(\! -\boldsymbol{\bar{\mathrm{D}}}_{1} t \big)\big\rVert_{\mathrm{F}} < \infty \;.
\label{eq:stabilityconditionII}%
\end{equation}
Since $\boldsymbol{\bar{\mathrm{D}}}_{1}$ is nilpotent, the matrix exponential 
$\exp(\! -\boldsymbol{\bar{\mathrm{D}}}_{1} t)$ is a finite-order polynomial in $-\boldsymbol{\bar{\mathrm{D}}}_{1} t$. 
With special exceptions, which constitute a null set among the set of all possible realizations of $\boldsymbol{\bar{\mathrm{D}}}$, $\exp(\! -\boldsymbol{\bar{\mathrm{D}}}_{1} t)$ can therefore be neglected in Eq.\ \eqref{eq:stabilityconditionII}.

Furthermore, we can take advantage of the fact that the Frobenius norm is invariant under unitary transformations and replace 
$\boldsymbol{\bar{\mathrm{D}}}_{0}$ by a corresponding diagonal matrix $\boldsymbol{\Lambda}=\Diag(\lambda_{1},\dotsc,\lambda_{n})$, where $\lambda_{1},\dotsc,\lambda_{n}$ are the eigenvalues of $\boldsymbol{\bar{\mathrm{D}}}_{0}$ and thus of $\boldsymbol{\bar{\mathrm{D}}}$. 
The stability criterion \eqref{eq:stabilityconditionII} then becomes 
\begin{equation}%
\lim_{t\to\infty} \big\lVert \exp(\! -\boldsymbol{\Lambda} t) \big\rVert_{\mathrm{F}} < \infty \;.
\label{eq:stabilityconditionIII}%
\end{equation}
This means that the mixture is stable, if the real parts of all eigenvalues of $\boldsymbol{\bar{\mathrm{D}}}$ are nonnegative, i.e., 
\begin{equation}%
\Real(\lambda_{i})\geqslant 0 \qquad \forall\, i\in\{1,\dotsc,n\} \;.
\label{eq:stabilityconditionIVb}%
\end{equation}
When one of the eigenvalues of $\boldsymbol{\bar{\mathrm{D}}}$ has a negative real part, the mixture is unstable.

\clearpage
\bibliographystyle{apsrev4-1}
\bibliography{References}

\end{document}